\documentclass[showpacs,aps,pra,superscriptaddress,longbibliography]{revtex4-2}
\usepackage{geometry}
\usepackage{amsfonts}
\usepackage{graphicx}
\usepackage{breakcites}
\usepackage[normalem]{ulem}
\usepackage{amsmath}
\usepackage{hyperref}
\usepackage{amssymb}
\usepackage{cleveref}
\usepackage{mathtools}
\usepackage{physics}
\usepackage{color}
\usepackage{ulem}
\usepackage[section]{placeins}
\usepackage[utf8]{inputenc}
\usepackage{comment}

\begin{document}
\title{Path-integral approach to the thermodynamics of bosons with memory: Density and correlation functions.}

\author{T. Ichmoukhamedov}
\author{J. Tempere}
\affiliation{TQC, Departement Fysica, Universiteit Antwerpen, Universiteitsplein 1, 2610 Antwerpen, Belgium}

\date{\today}

\begin{abstract}
Expanding upon previous work, using the path-integral formalism we derive expressions for the one-particle reduced density matrix and the two-point correlation function for a quadratic system of bosons that interact through a general class of memory kernels. 
The results are applied to study the density, condensate fraction and pair correlation function of trapped bosons harmonically coupled to external distinguishable masses.
\end{abstract}


\maketitle

\section{INTRODUCTION}

Retardation potentials describe interactions that depend not on the simultaneous position of the interacting particles, but on their positions at different times \cite{feynman1998statistical}. An example is the phonon-mediated interaction between electrons in a polar or ionic crystal. The phonons propagate slowly in comparison to the electrons. When the phononic degrees of freedom are integrated out, one is left with a retardation potential telling us that an electron feels the effect of another electron at an earlier time \cite{FeynmanPolaron}. Retardation potentials are not restricted to condensed matter systems: in electromagnetism, the Li\'enard-Wiechert potential between charges results from integrating out the photons. In theories of quantum dissipation, the effect of a (bosonic) bath can often be integrated out resulting in influence phases with retardation \cite{FeynmanVernon}. When the bath consists of an (infinite number of) harmonic oscillators, also the retarded potential is harmonic \cite{CaldeiraLeggett}.

Applications of retardation potentials such as the path-integral treatment of the polaron \cite{FeynmanPolaron} consider only one particle interacting with itself at an earlier time. In more advanced applications, one can consider more general forms of retardation potentials for a single particle \cite{Rosenfelder2001}, or consider many-body retardation effects such as encountered in variational approaches for many-polaron systems \cite{Klimin2004,Verbist1992,Casteels2013}. More recently, systems of distinguishable particles coupled to environments that could be integrated out, have also been the subject of various studies in the context of the thermodynamics of open quantum systems \cite{Ingold2009,Hasegawa1,Hasegawa2,Ingold2012,Ingold2014}.

However, for a system of many identical particles, the required symmetrization of the many-body density matrix complicates analytic calculations. In the context of non-retarded potentials, this symmetrization can be tackled by re-expressing the sum over particle permutations as a sum over cyclic decompositions of these permutations \cite{feynman1998statistical}. In the grand canonical ensemble the sum over all cyclic decompositions does not contain any constraint. However, in the canonical ensemble the condition of a fixed particle number results in a constraint on the sum over cyclic decompositions, inhibiting direct computation of the sum. In the context of the path-integral formalism, this problem has been studied for a system of coupled bosonic oscillators by Brosens \textit{et al.} \cite{Brosens1997a,Brosens1997b}, and some applications and extensions of this approach can be found in \cite{Brosens1998application,Tempere2000,Klimin2004}. In a recent work of the present authors \cite{paper1}, the work of Brosens \textit{et al.} was extended to retardation potentials. In particular, we considered a general class of quadratic many-body systems with retardation, as described by the action functional (in units of $\hbar=1$):
\begin{align}
S^{(N)}[\overline{\mathbf{r}}, x , y , \overline{\boldsymbol{\kappa}}]= &  \frac{m}{2} \sum_{i}^N \int_{0}^{ \beta}  \dot{\mathbf{r}}_i(\tau)^2 d\tau +  \frac{m}{2} \sum_{i}^N \int_0^{ \beta} d\tau \int_0^{\beta} d\sigma x(\tau - \sigma)  \mathbf{r_i}(\tau) \cdot \mathbf{r_i}(\sigma) \nonumber \\
 + \frac{m}{2N} & \sum_{i, j} \int_0^{ \beta} d\tau \int_0^{\beta} d\sigma \left[y(\tau-\sigma)-x(\tau-\sigma) \right]  \mathbf{r_i}(\tau) \cdot \mathbf{r_j}(\sigma) -m \sum_i^{N} \int_0^{\beta} d \tau \mathbf{r}_i (\tau) \cdot \boldsymbol{\kappa}_i(\tau).
\label{action1}
\end{align}
The action functional (\ref{action1}) describes $N$ particles with mass $m$ at an inverse temperature $\beta$, driven by source functions $\overline{\boldsymbol{\kappa}}$ and interacting between themselves through memory kernels $x$ and $y$. 
When the memory kernels are set equal to delta functions, the action functional reduces to the case without retardation, studied in \cite{Brosens1997a,Brosens1997b}, but more general choices can represent the effect of an environment that has been integrated out \cite{feynman1998statistical}. 
We will treat the memory kernels as generally as possible aside from the minimal assumptions of symmetry $x(\tau)=x(-\tau)$ and $\beta$-periodicity $x(\beta-\tau)=x(\tau)$ for both of the memory kernels, which allows to represent them as the Fourier series $x(\tau)=\sum_{n=-\infty}^{\infty} x_n e^{i\nu_n \tau}$, with $\nu_n= 2\pi n / \beta $ the bosonic Matsubara frequency. In addition we will assume that $x_0$ and $y_0$ are strictly non-zero, although this restriction can be omitted by introducing a finite system volume. 

In our above-mentioned previous work \cite{paper1}, the partition sum and some derived thermodynamic quantities such as the internal energy and the specific heat were calculated. In essence, we computed the trace of the density matrix. However, in order to use many-body systems with retardation as variational model systems, it is necessary to also know the one-and two-point correlation functions. These quantities give access to expectation values of single-particle operators (such as the density) and of two-body operators (such as the pair correlation function). The goal of the current paper is to derive the one-particle reduced density matrix and the two-point correlation function. As an example, we then apply these results to an open quantum system of bosons coupled to a model environment of distinguishable masses.

In Sec.~\ref{review}, a short review of previously obtained results is presented, and the path-integral definitions of the one-particle reduced density matrix and the two-point correlation function are given. We calculate expressions for the former in Sec.~\ref{section2}, and for the latter in Sec.~\ref{section3}. The obtained result for the two-point correlation function provides a generalization of expressions found in \cite{Brosens1997b} to systems with memory. Our result for the one-particle reduced density matrix allows for the computation of the effective states and occupation numbers of the bosons. In Sec.~\ref{section4} we apply these results to further explore the simplified model of bosons in an environment introduced in \cite{paper1} and in particular focus on the behavior of density and condensate fraction. Conclusions are drawn in Sec.~\ref{sec:conclusions}.

\section{Quadratic many-body systems with memory} \label{review}

In \cite{paper1} the distinguishable particle propagator corresponding to the action functional (\ref{action1}) was shown to be given by:
\begin{equation}
K_N[x, y, \overline{\boldsymbol{\kappa}} ] \left( \overline{\mathbf{r}}_T  , \beta |  \overline{\mathbf{r}}_0, 0 \right) = \frac{ K[y, \sqrt{N} \mathbf{K} ] (\sqrt{N} \mathbf{R}_T , \beta | \sqrt{N} \mathbf{R}_0, 0 )}{K[x , \sqrt{N} \mathbf{K} ] (\sqrt{N} \mathbf{R}_T , \beta | \sqrt{N} \mathbf{R}_0, 0 )} \prod_{j=1}^{N} K[x , \boldsymbol{\kappa}_j] (\mathbf{r}_{j,T}, \beta | \mathbf{r}_{j,0}, 0 ),
\label{propagatorfactorization1}
\end{equation}
where $\mathbf{R}=\frac{1}{N} \sum_i \mathbf{r}_i $ is the center of mass coordinate and $\mathbf{K}=\frac{1}{N} \sum_i \boldsymbol{\kappa}_i$ is the center of mass source term. 
The notation with square brackets indicates the dependence of the propagator on the system parameters, i.e. the memory kernels $x$ and $y$ and the set of sources $\overline{\boldsymbol{\kappa}}=\{\boldsymbol{\kappa}_j\}$. The sets of initial and final positions are denoted by $\overline{\mathbf{r}}_0=\{\mathbf{r}_{j,0}\}$ and $\overline{\mathbf{r}}_T=\{\mathbf{r}_{j,T}\}$, respectively.

The propagators on the RHS of (\ref{propagatorfactorization1}), $K[x , \boldsymbol{\kappa}_j] (\mathbf{r}_{j,T}, \beta | \mathbf{r}_{j,0}, 0 )$, are the single-particle propagators corresponding to the $N=1$ limit of (\ref{action1}), and only depend on a single memory kernel:
\begin{align}
K[x , \boldsymbol{\kappa}] (\mathbf{r}_T, \beta | \mathbf{r}_0, 0 ) &=   \mathcal{A}^{d}  \exp \left[ -\frac{m}{2 \beta} A_x (\mathbf{r}_T - \mathbf{r}_0)^2 - \frac{m}{2\beta} \frac{1}{\Delta_{x}(0)} (\mathbf{r}_T + \mathbf{r}_0)^2 
\nonumber \right. \\
&+  \frac{2m}{\beta} \frac{1}{\Delta_{x}(0)}    \sum_{n } \frac{\boldsymbol{\kappa}_n  }{\nu_n^2 + \beta x_n} \cdot   (\mathbf{r}_T + \mathbf{r}_0) - \frac{2 m }{\beta} \left( \frac{\beta}{2}    \sum_{n \neq 0}  \frac{ i \nu_n}{   \nu_n^2 + \beta x_n }  \boldsymbol{\kappa}_n \right)  \cdot  (\mathbf{r}_T - \mathbf{r}_0)  \nonumber  \\
&\left. -\frac{2 m }{\beta} \frac{1}{\Delta_{x}(0)} \left( \sum_n \frac{\boldsymbol{\kappa}_{n}}{\nu_n^2 + \beta x_n}  \right)^2 + \frac{2m}{\beta} \left( \frac{\beta^2}{4}   \sum_n \frac{\boldsymbol{\kappa}_n \cdot \boldsymbol{\kappa}_{-n}}{\nu_n^2 + \beta x_n} \right) \right],
\label{SingleParticlePropagator}
\end{align}
with
\begin{align}
\mathcal{A}&= \left( \frac{m}{2 \pi \beta} \right)^{1/2} \left(  \frac{4}{\beta^3 x_0 \Delta_x(0)} \right)^{1/2} \frac{1}{ {\displaystyle \prod_{k=1}  \left(  1 + \frac{\beta x_k}{\nu_k^2}\right) } }.
\label{mathcalAmain}
\end{align}
Note that we will also be using the shorthand notation $K[x ]=K[x , \boldsymbol{0}]$ further on, corresponding to setting all $\boldsymbol{\kappa}_n$ to zero in expression (\ref{SingleParticlePropagator}), which leaves just the first two terms in the exponent. The two dimensionless functionals $A_x$ and $\Delta_x(\tau)$ appearing in the propagator are defined as:
\begin{align}
&A_x= \sum_{n=-\infty}^{\infty} \frac{\beta x_n}{\nu_n^2 + \beta x_n}, \label{Axn} \\
&\Delta_x(\tau)=  \frac{4}{\beta^2} \sum_{n= -\infty}^{\infty} \frac{e^{i \nu_n \tau}}{\nu_n^2 + \beta x_n}. \label{Dxn}
\end{align}
We will use the shorthand notation $\Delta_x=\Delta_x(0)$ and in general we will assume $A_x >0$ and $\Delta_x >0$ in this paper to restrict the memory kernels to produce bounded propagators as a function of the end-points (\ref{SingleParticlePropagator}).

Let us write the partition function of a system of bosons described by (\ref{action1}) as $\mathcal{Z}[\boldsymbol{\overline{\kappa}}](N)$, where the dependence on the source functions is explicitly highlighted in the functional. 
In \cite{paper1} the partition function $\mathcal{Z}(N)=\mathcal{Z}[\boldsymbol{0}](N)$ for this system without source terms, $\boldsymbol{\overline{\kappa}}=\mathbf{0}$, was calculated and applied to study the specific heat of a model of an open quantum system.  
The goal of this paper is to expand upon this calculation and derive expressions for the one-particle reduced density matrix (from which the one-point correlation function readily follows) and the two-point correlation function. 
The one-particle reduced density matrix is computed in a similar way as the partition function in \cite{paper1}, but now the integration variable $\mathbf{r}_1$ is removed from the integral resulting in an $N-1$ dimensional integral over $\mathbf{\tilde{r}}=\lbrace \mathbf{r}_2,...,\mathbf{r}_N \rbrace$, and in the boundary points of the path integral $\mathbf{r}_1$ is replaced by respectively $\mathbf{r}$ and $\mathbf{r}'$:

\begin{align}
\rho_1(\mathbf{r}'|\mathbf{r})&=\frac{1}{\mathcal{Z}(N)}  \frac{1}{N!}   \sum_{P}  
\int d\mathbf{\tilde{r}}
\int_{\lbrace \mathbf{r}, \mathbf{r}_2, ..., \mathbf{r}_N \rbrace,0}^{P[ \lbrace \mathbf{r}', \mathbf{r}_2, ..., \mathbf{r}_N \rbrace], \beta } \mathcal{D} 
\overline{\mathbf{r}}''  e^{-S^{(N)}[\overline{\mathbf{r}}'', x , y , \boldsymbol{0}]} . \label{1RDM}
\end{align}
For the computation of the two-point correlation function, the structure of the path-integral is somewhat simpler as all variables $\overline{\mathbf{r}}=\lbrace \mathbf{r}_1, \mathbf{r}_2,... \mathbf{r}_N \rbrace$ are treated on the same footing, but an additional weighing factor containing two different variables $\mathbf{r}_1$ and $\mathbf{r}_2$ (identical for any $i \neq j$) appears:
\begin{align}
\expval{ e^{i \mathbf{q} \cdot( \mathbf{r}_1(\tau) - \mathbf{r}_2(\sigma) )} }_I&=\frac{1}{\mathcal{Z}(N)}  \frac{1}{N!} \sum_{P}  
\int d\overline{\mathbf{r}}
\int_{\overline{\mathbf{r}},0}^{P[\overline{\mathbf{r}}], \beta } \mathcal{D} 
\overline{\mathbf{r}}' ~e^{i \mathbf{q} \cdot( \mathbf{r}_i(\tau) - \mathbf{r}_j(\sigma) )}  e^{-S^{(N)}[\overline{\mathbf{r}}', x , y , \boldsymbol{0}]} . \label{twopoint1}
\end{align}
In the next two sections, we perform the many-body path integrations in expressions (\ref{1RDM}) and (\ref{twopoint1}).

\section{One-particle reduced density matrix} \label{section2}

The derivation of the one-particle reduced density matrix can be summarized as a modification of the derivation of the partition function in \cite{paper1} to account for the unequal treatment of the variables $\mathbf{r}$ and $\mathbf{r}'$ in comparison to the other variables $ \mathbf{\tilde{r}}= \lbrace \mathbf{r}_2,...,\mathbf{r}_N \rbrace$. 
The first step is to note that for any fixed permutation, the path integral in (\ref{1RDM}) is given by the propagator (\ref{propagatorfactorization1}), where the center of mass part is unaffected by the permutations while the rest of the end points have to be permuted accordingly. 
The center of mass can now be isolated in the same spirit as in \cite{Brosens1997a,paper1} using a modified variable $\mathbf{\tilde{R}}=\frac{1}{N} \sum_{j=2}^{N} \mathbf{r}_j$ that only contains integration variables. This allows us to write:

\begin{equation}
\rho_1(\mathbf{r}'|\mathbf{r})  = \frac{1}{(2 \pi)^d} \frac{1}{\mathcal{Z}(N)} \int d\mathbf{k} ~ \mathcal{P}_{\tilde{R}} (N,\mathbf{k})  \mathcal{P}_r(N,\mathbf{k}) , \label{kintegralrho}
\end{equation}
where the center of mass contribution is the Gaussian integral:

\begin{equation}
\mathcal{P}_{\tilde{R}} (N,\mathbf{k})=  \int d\mathbf{\tilde{R}} ~ e^{i \mathbf{k} \cdot \mathbf{\tilde{R}}} \frac{ K[y] \left( \sqrt{N} \mathbf{\tilde{R}}+ \frac{1}{\sqrt{N}} \mathbf{r} , \beta | \sqrt{N} \mathbf{\tilde{R}}+ \frac{1}{\sqrt{N}} \mathbf{r}', 0 \right)}{ K[x] \left( \sqrt{N} \mathbf{\tilde{R}}+ \frac{1}{\sqrt{N}} \mathbf{r} , \beta | \sqrt{N} \mathbf{\tilde{R}}+ \frac{1}{\sqrt{N}} \mathbf{r}', 0 \right)}, \label{CMpartitionfunctionrho}
\end{equation}
which is readily computed if we assume that $\Delta_x > \Delta_y$. 
The remaining factor contains the permutations:

 \begin{equation}
 \mathcal{P}_r(N,\mathbf{k})= \frac{1}{N!} \sum_P    \int d\mathbf{\tilde{r}}  K[x ] \left(  \mathbf{r}_{P(N)}, \beta | \mathbf{r}_N, 0 \right) ... K[x ] \left( \mathbf{r}', \beta |  \mathbf{r}_j, 0 \right)... K[x ] \left(  \mathbf{r}_{P(1)}, \beta | \mathbf{r}, 0 \right)  e^{ - i \mathbf{k} \cdot  \sum_{j=2}^{N} \mathbf{r}_j /N} ,
\label{Ppartitionfunction}
\end{equation}
where $P(n)$ represents the element that ends up at the position of $n$ after the permutation on the ordered set $\lbrace 1,2,...,N \rbrace$. 
Expression (\ref{Ppartitionfunction}) illustrates how the end-point $\mathbf{r}'$ of the one-particle reduced density matrix is permuted to some position $j$, while the initial point $\mathbf{r}$ remains in place and gets coupled with element $P(1)$.

\begin{figure}[!htb]
\includegraphics[width=0.9\columnwidth]{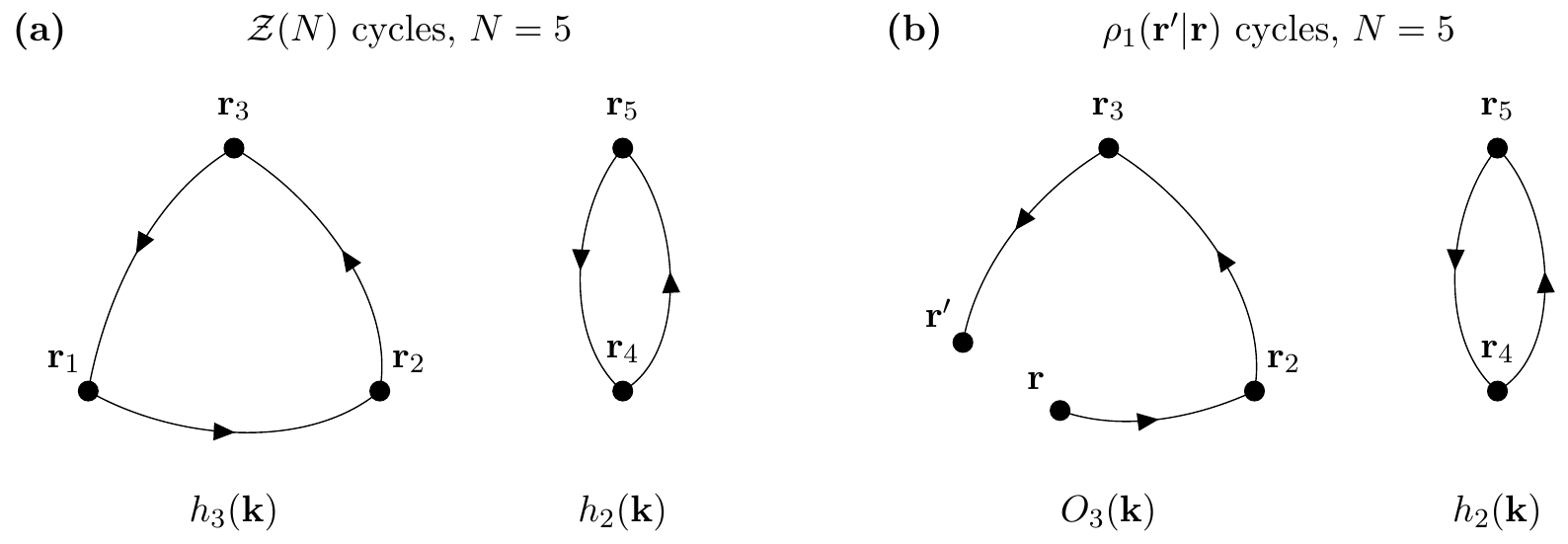}
\caption{Depiction of the different types of cycles for $N=5$ where the arrows represent the single-particle propagators (\ref{SingleParticlePropagator}). The cyclic decomposition of the partition function in the absence of source terms \cite{paper1} is illustrated in (a) where the closed cycles contribute a factor $h_{\ell}(\mathbf{k})$ given by (\ref{hqmain}). The modified decomposition of the 1RDM $\rho_1(\mathbf{r}'|\mathbf{r})$ is shown in (b), where one cycle is now opened up contributing a factor $O_\ell(\mathbf{k})$ in (\ref{Ppartitionfunction2}) while the remaining points are still partitioned in terms of closed cycles.} 
\label{Figure1}
\end{figure}

We now follow the standard approach \cite{feynman1998statistical,Brosens1997a,paper1} to decompose the summation of permutations in (\ref{Ppartitionfunction}) in terms of its cyclic decomposition, an argument which will require some modification for the calculation of $\rho_1$.
It is easy to see that most cycles will be completely unaffected by the presence of $\mathbf{r}$ and $\mathbf{r}'$ and will yield exactly the same contribution $h_{\ell}(\mathbf{k})$ as computed for the partition function in \cite{paper1}. In fact, there will be only one modified permutation chain starting at $\mathbf{r} \rightarrow \mathbf{r}_{P(1)} \rightarrow ... $ that has to end at $\mathbf{r}'$. 
As illustrated in Fig.~\ref{Figure1}, this chain can be thought of as an open cycle with length $\ell$ the contribution of which we will call $O_\ell (\mathbf{k})$.
The summation in (\ref{Ppartitionfunction}) can therefore be written as the summation over all the possible open cycles $O_\ell (\mathbf{k})$ multiplied with the cyclic decomposition on the remaining $N-\ell$ points (and taking the combinatorics into account): 
 \begin{equation}
 \mathcal{P}_r  (N,\mathbf{k})= \frac{1}{N}  \sum_{\ell=1}^{N}  O_\ell  (\mathbf{k})  \sum_{M_1,...,M_{N-\ell}}^{*} \prod_{n=1}^{N-\ell} \frac{h_n(\mathbf{k})^{M_n}}{n^{M_n} M_n!}. \label{Ppartitionfunction2}
\end{equation}
Here, the ordinary closed cycles $h_n(\mathbf{k})$ were computed in \cite{paper1}:
\begin{equation}
h_\ell (\mathbf{k})= Q_x^{\ell d} \frac{1}{ \left[  2 \left| \sinh( \frac{ \ell }{2}  \textrm{arccosh}\left[ \frac{A_x \Delta_{x} + 1}{A_x \Delta_{x} - 1} \right] )  \right|  \right]^d}  \exp \left( -\frac{ \ell k^2 \beta }{ 8 N^2 m} \Delta_{x}   \right) \label{hqmain}, 
\end{equation}
with:
\begin{equation}
Q_x=\frac{1}{ \prod_{k=1}  \left(  1 + \frac{\beta x_k}{\nu_k^2}\right) }   \left( \frac{1}{\beta^3 x_0}  \frac{ 4}{  \left|  A_x \Delta_{x} - 1  \right| } \right)^{1/2},
\label{Qxn}
\end{equation}
and the open cycles are given by:
\begin{equation}
O_\ell (\mathbf{k}) =  \int d\mathbf{r}_2  ... \int d\mathbf{r}_\ell ~  K[x ] ( \mathbf{r}', \beta | \mathbf{r}_\ell, 0 ) ... K[x ] ( \mathbf{r}_3, \beta | \mathbf{r}_2 , 0 )   K[x ] ( \mathbf{r}_2, \beta | \mathbf{r} , 0 ) e^{ - \frac{i\mathbf{k}  }{N}   \cdot  \sum_{j=2}^{\ell} \mathbf{r}_j}. \label{O1}
\end{equation}
This integral is computed in Appendix \ref{appendixA} and is shown to be equal to:
\begin{align}
O_\ell (\mathbf{k}) = & \mathcal{A}^{\ell d}    \left( \frac{\pi^{\ell-1}}{ \left[ \frac{m}{2\beta} \left( A_x - \frac{1}{\Delta_x} \right) \right]^{\ell-1} U_{\ell-1}( \zeta) } \right)^{d/2}   \exp \left( - \frac{ \beta k^2}{8 N^2 m} \Delta_x \left[  \ell - \sqrt{ A_x \Delta_x}  \tanh( \frac{\ell}{2} \textrm{arccosh}(\zeta) )  \right]   \right. \nonumber \\
&  - \frac{i \mathbf{k} }{2N} \cdot (\mathbf{r}+\mathbf{r}')\left[  \sqrt{ A_x \Delta_x} \tanh( \frac{\ell}{2} \textrm{arccosh}(\zeta) )  -1 \right] \nonumber        \\
&  \left. - \frac{m}{\beta} \sqrt{\frac{A_x}{\Delta_x}}  \tanh( \frac{\ell}{2} \textrm{arccosh}(\zeta) )   (\mathbf{r}^2+ \mathbf{r}^{\prime 2}) -  \frac{m}{\beta} \sqrt{\frac{A_x}{\Delta_x}}  \frac{1}{\sinh(\ell~\textrm{arccosh}\left(\zeta\right) ) } (\mathbf{r} - \mathbf{r}')^2  \right) 
\label{O2},
\end{align}
where 
\begin{equation}
\zeta= \frac{A_x \Delta_x +1}{A_x \Delta_x -1},    
\end{equation}
and $U_{\ell-1}$ is a Chebyshev polynomial of the second kind defined in expression (\ref{Chebyshev}) of Appendix \ref{appendixA} for $|\zeta|>1$.

The closed cycles $h_n(\mathbf{k})$ (\ref{hqmain}) can be now substituted in (\ref{Ppartitionfunction2}) after which the Fourier integral in (\ref{kintegralrho}) can be readily performed. 
The final expression for the one particle reduced density matrix is found as:

\begin{align}
\rho_1(\mathbf{r}'|\mathbf{r})  =   &  \left( \frac{2m}{\pi \beta} \right)^{d  /2}    \frac{1}{N} \sum_{\ell=1}^{N}  \frac{ \mathbb{Z}(N-\ell) }{\mathbb{Z}(N)  } \left(     \frac{1 }{   \left[  \sqrt{\frac{\Delta_x}{A_x}}\coth( \frac{\ell}{2} \textrm{arccosh}(\zeta) )  -  \frac{1}{N} (\Delta_x-\Delta_y)  \right]}\right)^{d/2}  \nonumber  \\
& \frac{ 1 }{    \left| 2\sinh( \frac{\ell}{2} \textrm{arccosh}(\zeta) ) \right|^{d}}  \exp \left( -\frac{m}{2\beta} \frac{ \frac{1}{N} (\Delta_x - \Delta_y)  \sqrt{\frac{A_x}{\Delta_x}}   \tanh( \frac{\ell}{2} \textrm{arccosh}(\zeta) ) }{ \sqrt{\frac{\Delta_x}{A_x}} \coth( \frac{\ell}{2} \textrm{arccosh}(\zeta) )- \frac{1}{N}(\Delta_x - \Delta_y)  } (\mathbf{r}+\mathbf{r}')^2  \right) \nonumber \\
& \exp \left( - \frac{m}{\beta} \sqrt{\frac{A_x}{\Delta_x}}  \tanh( \frac{\ell}{2} \textrm{arccosh}(\zeta) )   (\mathbf{r}^2+ \mathbf{r}^{\prime 2})  \right) \\
& \exp \left(- \frac{m}{2\beta} \left[ \frac{1}{N} (A_y-A_x) +   \sqrt{\frac{A_x}{\Delta_x}}  \frac{2}{\sinh(\ell~\textrm{arccosh}\left(\zeta\right) ) } \right] (\mathbf{r} - \mathbf{r}')^2  \right) \label{1RDMfinal}.
\end{align}
Here, $\mathbb{Z}(N)$ can be found as the solution to the recurrence relation (with $\mathbb{Z}(0)=1$) studied in \cite{paper1}:

\begin{equation}
\mathbb{Z}(N) =  \frac{1}{N} \sum_{k=0}^{N-1} \mathbb{Z}(k) \frac{ 1 }{    \left| 2\sinh( \frac{(N-k)}{2} \textrm{arccosh}(\zeta) ) \right|^{d}} .
\label{recurrence1}
\end{equation}
Note that by relabeling $k=N-\ell$ it follows from (\ref{recurrence1}) that for any $N$:
\begin{equation}
\frac{1}{N}     \sum_{\ell=1}^{N}  \frac{ \mathbb{Z}(N-\ell) }{\mathbb{Z}(N)  }  \frac{ 1 }{    \left| 2\sinh( \frac{\ell}{2} \textrm{arccosh}(\zeta) ) \right|^{d}}  = 1,
\end{equation}
which guarantees that the one-particle reduced density matrix (\ref{1RDMfinal}) is always normalized.

The diagonal of $\rho_1$ in position space can be readily taken and yields the average boson density. Note that here and in the rest of this paper the density will be normalized to $1$ rather than to $N$:

\begin{align}
n(\mathbf{r})=\rho_1(\mathbf{r}|\mathbf{r})  =   &  \left( \frac{2m}{\pi \beta} \right)^{d  /2}    \frac{1}{N} \sum_{\ell=1}^{N}  \frac{ \mathbb{Z}(N-\ell) }{\mathbb{Z}(N)  } \left(     \frac{1 }{   \left[  \sqrt{\frac{\Delta_x}{A_x}}\coth( \frac{\ell}{2} \textrm{arccosh}(\zeta) )  -  \frac{1}{N} (\Delta_x-\Delta_y)  \right]}\right)^{d/2}  \nonumber  \\
&\frac{ 1 }{    \left| 2\sinh( \frac{\ell}{2} \textrm{arccosh}(\zeta) ) \right|^{d}}   \exp \left( - \frac{2m}{\beta}  \frac{1}{ \sqrt{\frac{\Delta_x}{A_x}} \coth( \frac{\ell}{2} \textrm{arccosh}(\zeta) )- \frac{1}{N}(\Delta_x - \Delta_y)  }     \mathbf{r}^2  \right). 
\label{density1}
\end{align}
We can now consider a specific choice of memory kernels $x(\tau-\sigma)= \omega^2 \delta(\tau-\sigma)$ and $y(\tau-\sigma)=\Omega^2 \delta(\tau-\sigma)$ in which case the action functional (\ref{action1}) reduces to that of a system of harmonically trapped bosons all coupled by springs, as studied in \cite{Brosens1997b}. 
In this case the Matsubara summations (\ref{Axn}) and (\ref{Dxn}) can be computed to find $\Delta_x = \frac{2}{\beta \omega} \coth (\frac{\beta \omega}{2})$, $\Delta_y = \frac{2}{\beta \Omega} \coth (\frac{\beta \Omega}{2})$ and $A_x= \frac{\beta \omega}{2} \coth (\frac{\beta \omega}{2})$. In addition $\zeta= \cosh (\frac{\beta \omega}{2})$ which nicely cancels with the $\textrm{arccosh}$ function in the argument of the hyperbolic sine. 
After substitution in (\ref{density1}), the expression for the density in \cite{Brosens1997b} is retrieved exactly.

\section{Two-point correlation function}\label{section3}

The goal of this section is to find an expression for the partition function with two general non-zero source terms $\boldsymbol{\overline{\kappa}}_2=(\boldsymbol{\kappa_1},\boldsymbol{\kappa_2},0,...)$:

\begin{equation}
\mathcal{Z}[\boldsymbol{\overline{\kappa}}_2](N)=\frac{1}{N!}  \sum_{P}  
\int d\overline{\mathbf{r}}
\int_{\overline{\mathbf{r}},0}^{P[\overline{\mathbf{r}}], \beta } \mathcal{D} 
\overline{\mathbf{r}}' ~   e^{-S^{(N)}[\overline{\mathbf{r}}', x , y ,\boldsymbol{\overline{\kappa}}_2]} ,
\label{Zkappa2}
\end{equation}
which after division by $\mathcal{Z}(N)$ and setting $\boldsymbol{\kappa}_1(\tau')= \frac{i \mathbf{q}}{m} \delta(\tau-\tau')$ and $\boldsymbol{\kappa}_2(\sigma')= -\frac{i \mathbf{q}}{m} \delta(\sigma-\sigma')$  yields exactly the two point correlation function (\ref{twopoint1}).
First, the propagator (\ref{propagatorfactorization1}) is substituted in (\ref{Zkappa2}) where two source terms are set non-zero. Contrary to the approach in Section \ref{section2}, all of the variables $\mathbf{\overline{r}}$ are integrated out in (\ref{Zkappa2}). 
The center of mass can therefore be separated using the complete CM variables $\mathbf{R}=\frac{1}{N} \sum_{i=1}^{N} \mathbf{r}_i$ and $\mathbf{K}_2=\frac{1}{N}    ( \boldsymbol{\kappa}_1 + \boldsymbol{\kappa}_2) $ which allows to write:
\begin{equation}
\mathcal{Z}[\boldsymbol{\overline{\kappa}}_2](N) = \frac{1}{(2 \pi)^d} \int d\mathbf{k} ~ \mathcal{Z}_R [\boldsymbol{\overline{\kappa}}_2](N,\mathbf{k})  \mathcal{Z}_r[\boldsymbol{\overline{\kappa}}_2](N,\mathbf{k}) , \label{kintegral}
\end{equation}
where:
\begin{equation}
\mathcal{Z}_R [\boldsymbol{\overline{\kappa}}_2] (N,\mathbf{k})=  \int d\mathbf{R} ~\frac{ K[y, \sqrt{N} \mathbf{K}_2 ] (\sqrt{N} \mathbf{R} , \beta | \sqrt{N} \mathbf{R} , 0 )}{K[x, \sqrt{N} \mathbf{K}_2 ] (\sqrt{N} \mathbf{R}  , \beta | \sqrt{N} \mathbf{R} , 0 )}, \label{CMpartitionfunction}
\end{equation}
and:
 \begin{align}
 \mathcal{Z}_r [\boldsymbol{\overline{\kappa}}_2] (N,\mathbf{k})  = \frac{1}{N!} &\sum_P  \int d\overline{\mathbf{r}} ~ e^{ - \frac{i \mathbf{k}}{N} \cdot \sum_{j=1}^{N} \mathbf{r}_j } \prod_{j=3}^{N} K[x ] ( P\mathbf{r}_j, \beta | \mathbf{r}_j, 0 )     \nonumber \\
 &  K[x,\boldsymbol{\kappa_2} ] ( P\mathbf{r}_2, \beta | \mathbf{r}_2, 0 ) K[x,\boldsymbol{\kappa_1} ] ( P\mathbf{r}_1, \beta | \mathbf{r}_1, 0 ) .\label{rpartitionfunction}
\end{align}
In expression (\ref{rpartitionfunction}) all source functions are set to zero except for the two sources $\boldsymbol{\kappa}_1$ and $\boldsymbol{\kappa}_2$ corresponding to the propagators starting in initial points $\mathbf{r}_1$ and $\mathbf{r}_2$. 
Making use of expression (\ref{SingleParticlePropagator}) for the propagator, (\ref{rpartitionfunction}) can also be written as:
 \begin{align}
 \mathcal{Z}_r [\boldsymbol{\overline{\kappa}}_2] (N,\mathbf{k})  = \frac{1}{N!} & e^{ -S^{(1)}_{\textrm{cl}} \left[ x, \boldsymbol{\kappa}_1\right] \left( 0,0 \right)-S^{(1)}_{\textrm{cl}} \left[ x, \boldsymbol{\kappa}_2\right] \left( 0,0 \right)} \sum_P  \int d\overline{\mathbf{r}} ~ e^{ - \frac{i \mathbf{k}}{N} \cdot \sum_{j=1}^{N} \mathbf{r}_j }      \nonumber \\
 &  e^{\mathbf{\tilde{a}_1}(P\mathbf{r}_1+\mathbf{r}_1) +\mathbf{\tilde{b}_1} (P\mathbf{r}_1-\mathbf{r}_1)} e^{\mathbf{\tilde{a}_2} (P\mathbf{r}_2+\mathbf{r}_2) +\mathbf{\tilde{b}_2}(P\mathbf{r}_2-\mathbf{r}_2)} \prod_{j=1}^{N} K[x ] ( P\mathbf{r}_j, \beta | \mathbf{r}_j, 0 ) \label{rpartitionfunction2},
\end{align}
where the path-independent contribution of the propagators was taken out of the integral by using the notation:

\begin{equation}
S^{(1)}_{\textrm{cl}} \left[ x, \boldsymbol{\kappa}\right] \left( 0,0 \right)= \frac{2 m }{\beta} \frac{1}{\Delta_{x}(0)} \left( \sum_n \frac{\boldsymbol{\kappa}_{n}}{\nu_n^2 + \beta x_n}  \right)^2 - \frac{2m}{\beta} \left( \frac{\beta^2}{4}   \sum_n \frac{\boldsymbol{\kappa}_n \cdot \boldsymbol{\kappa}_{-n}}{\nu_n^2 + \beta x_n} \right).
\end{equation}
In addition we define the short-hand notation for the linear terms in the exponent corresponding to the two source terms $s=1,2$:
\begin{align}
&\mathbf{\tilde{a}_s} =\frac{2m}{\beta} \frac{1}{\Delta_{x}(0)}    \sum_{n } \frac{\boldsymbol{\kappa}_{\mathbf{s},n} }{\nu_n^2 + \beta x_n}, \hspace{20pt} \mathbf{\tilde{b}_s} =- \frac{2 m }{\beta} \left( \frac{\beta}{2}    \sum_{n \neq 0}  \frac{ i \nu_n}{   \nu_n^2 + \beta x_n }  \boldsymbol{\kappa}_{\mathbf{s},n} \right).
\end{align}
The approach to compute (\ref{rpartitionfunction2}) is once again to decompose the permutation in terms of cycles just as in \cite{paper1,Brosens1997b} or Section (\ref{section2}), with some modifications. 

Consider any general permutation on $N$ points out of the summation in (\ref{rpartitionfunction2}). 
For $2 \leq \ell \leq N$ there will be $N-\ell$ points that form cycles that do not pass through either $\mathbf{r}_1$ or $\mathbf{r}_2$ and yield the ordinary closed cycle contributions $h_n(\mathbf{k})$ as given in (\ref{hqmain}). To account for the cycle(s) on the remaining $\ell$ points, the set of all permutations has to be partitioned into two classes. 
In one class of permutations, the points $\mathbf{r}_1$ or $\mathbf{r}_2$ will lie in two disjoint cycles of respectively length $\ell$ and $\ell-j$ with $1 \leq j \leq \ell -1$. 
In this case the contribution of the two cycles can therefore be written as a product $ H_{j}^{(1)} (\mathbf{k})  H_{\ell-j}^{(2)}(\mathbf{k})$. In the second class of permutations, those points will be in the same cycle of length $\ell$ and yield a single contribution $\chi_{\ell}(\mathbf{k},j')$ which also depends on the distance between the two points $j'$ within this cycle, with $1 \leq j' \leq \ell -1$. 
After taking the combinatorics into account, this reasoning is written down as:
\begin{align}
 \mathcal{Z}_r [\boldsymbol{\overline{\kappa}}_2] (N,\mathbf{k}) =& \frac{e^{ -S^{(1)}_{\textrm{cl}} \left[ x, \boldsymbol{\kappa}_1\right] \left( 0,0 \right)-S^{(1)}_{\textrm{cl}} \left[ x, \boldsymbol{\kappa}_2\right] \left( 0,0 \right)}}{N(N-1)} \left[ \sum_{\ell=2}^{N} \sum_{j=1}^{\ell-1} H^{(1)}_{j}  (\mathbf{k})  H^{(2)}_{\ell-j}  (\mathbf{k})\sum_{M_1,...,M_{N-\ell}}^{*} \prod_{n=1}^{N-\ell} \frac{h_n(\mathbf{k})^{M_n}}{n^{M_n} M_n!} \right. \nonumber \\
 +& \left. \sum_{\ell=2}^{N} \sum_{j'=1}^{\ell-1} \chi_{\ell}(\mathbf{k},j')\sum_{M_1,...,M_{N-\ell}}^{*} \prod_{n=1}^{N-\ell} \frac{h_n(\mathbf{k})^{M_n}}{n^{M_n} M_n!} \right] \label{Ppartitionfunction3},
\end{align}
and after relabeling the integration variables we can write for $n=1,2$:
\begin{equation}
H^{(n)}_\ell (\mathbf{k}) = \int d\mathbf{r}_1  ... \int d\mathbf{r}_\ell ~ e^{\mathbf{\tilde{a}_n}(\mathbf{r}_2+\mathbf{r}_1) +\mathbf{\tilde{b}_n}(\mathbf{r}_2-\mathbf{r}_1)} K[x ] ( \mathbf{r}_1, \beta | \mathbf{r}_\ell, 0 ) ...  K[x ] ( \mathbf{r}_2, \beta | \mathbf{r}_1 0 ) e^{ - \frac{i\mathbf{k}  }{N}   \cdot  \sum_{j=1}^{\ell} \mathbf{r}_j}, \label{H1}
\end{equation}
and:
\begin{align}
\chi_{\ell}(\mathbf{k},j) = \int d\mathbf{r}_1  ... \int d\mathbf{r}_\ell ~ & e^{\mathbf{\tilde{a}_1}(\mathbf{r}_2+\mathbf{r}_1) +\mathbf{\tilde{b}_1}(\mathbf{r}_2-\mathbf{r}_1)} K[x ] ( \mathbf{r}_1, \beta | \mathbf{r}_\ell, 0 ) ...  K[x ] ( \mathbf{r}_2, \beta | \mathbf{r}_1 0 ) e^{ - \frac{i\mathbf{k}  }{N}   \cdot  \sum_{j=1}^{\ell} \mathbf{r}_j} \nonumber \\
&e^{\mathbf{\tilde{a}_2}(\mathbf{r}_{j+2}+\mathbf{r}_{j+1}) +\mathbf{\tilde{b}_2}(\mathbf{r}_{j+2}-\mathbf{r}_{j+1})} \label{chi1}.
\end{align}
In Appendix \ref{appendixB} expressions for both types of cycles are derived. 
If the notation from Appendix \ref{appendixB} is used, $a= \frac{m}{2\beta} A_x$, $b=\frac{m}{2\beta} \frac{1}{\Delta_x}$, to keep the expressions compact, the cycles can be written as:
\begin{align}
H^{(n)}_\ell (\mathbf{k})= & Q_x^{\ell d} \frac{1}{ \left[  2 \left|  \sinh( \frac{ \ell }{2}  \textrm{arccosh}\left( \zeta \right) )  \right| \right]^d}   \exp \left( - \frac{ \ell k^2}{16 N^2 b} -  \frac{ i \mathbf{k} \cdot \mathbf{\tilde{a}_n} }{4Nb } \right.  \nonumber \\
& \left.  +  \frac{  \mathbf{\tilde{a}_n} ^2 a - \mathbf{\tilde{b}_n} ^2 b  }{a-b} \frac{1}{4 \sqrt{ab}} \coth( \frac{\ell}{2} \textrm{arccosh}\left( \zeta \right) ) - \frac{1}{4} \frac{\mathbf{\tilde{a}_n} ^2-\mathbf{\tilde{b}_n} ^2}{a-b} \right), \label{Hresultmain} 
\end{align}
and:
\begin{align}
\chi_{\ell}(\mathbf{k},j)= & Q_x^{\ell d} \frac{1}{ \left[  2 \left|  \sinh( \frac{ \ell }{2}  \textrm{arccosh}\left( \zeta \right) )  \right| \right]^d}   \exp \left( \frac{-\ell k^2}{16N^2b} -  \frac{i \mathbf{k}}{4N b} \cdot  \left(\mathbf{\tilde{a}_1}+ \mathbf{\tilde{a}_2} \right)   \right. \nonumber \\
&+ \frac{1}{4} \left( \mathbf{\tilde{a}_1}^2+ \mathbf{\tilde{b}_1}^2+\mathbf{\tilde{a}_2}^2+\mathbf{\tilde{b}_2}^2 \right) D_0 + \frac{1}{4} \left( \mathbf{\tilde{a}_1}^2 - \mathbf{\tilde{b}_1}^2+\mathbf{\tilde{a}_2}^2 - \mathbf{\tilde{b}_2}^2 \right)  D_1  + \frac{1}{2} \left( \mathbf{\tilde{a}_1} \cdot \mathbf{\tilde{a}_2} + \mathbf{\tilde{b}_1} \cdot \mathbf{\tilde{b}_2} \right) D_j \nonumber \\
& \left. + \frac{1}{4} ( \mathbf{\tilde{a}_1} - \mathbf{\tilde{b}_1}) \cdot ( \mathbf{\tilde{a}_2} +\mathbf{\tilde{b}_2}) D_{j+1}+ \frac{1}{4}  (\mathbf{\tilde{a}_1} +\mathbf{\tilde{b}_1})\cdot (\mathbf{\tilde{a}_2}-\mathbf{\tilde{b}_2}) D_{j-1} \right),
\label{Chiresultmain}
\end{align}
where:
\begin{equation}
D_\ell(n)= \frac{1}{ 2\sqrt{ab}}  \frac{\cosh \left( \left(\frac{\ell}{2}-n \right) \textrm{arccosh}\left(\zeta \right) \right)}{\sinh \left( \frac{\ell}{2} ~ \textrm{arccosh}\left(\zeta \right) \right)}.
\end{equation}
All that remains now is to compute the Gaussian integral in the center of mass part (\ref{CMpartitionfunction}), and then combine the resulting expression with  (\ref{Ppartitionfunction3}) to compute the Fourier integral in (\ref{kintegral}). 
Although a rather lengthy calculation, it is reliant only on basic Gaussian integrals and goniometric identities, and we proceed to the final result:
\begin{align}
\frac{\mathcal{Z}[\boldsymbol{\overline{\kappa}}_2](N)}{\mathcal{Z}(N)}&= \frac{1}{N(N-1)} e^{-   \tilde{S}[  \boldsymbol{\overline{\kappa}}_2]}   \sum_{\ell=2}^{N} \frac{\mathbb{Z}(N-\ell) }{\mathbb{Z}(N )}  h_{\ell}(\mathbf{0}) Q_x^{ -\ell  d} \sum_{j=1}^{\ell-1} \left[   \frac{ h_{j}(\mathbf{0})  h_{\ell-j}(\mathbf{0})}{h_{\ell}(\mathbf{0}) }  \right. \nonumber \\
& \times \exp \left(  -  \frac{2m}{\beta}  \frac{1}{\sqrt{A_x \Delta_x}} \left[    \mathcal{J}[\boldsymbol{\kappa}_1, \boldsymbol{\kappa}_1]\coth( \frac{\ell}{2} \textrm{arccosh}\left( \zeta \right) )  +    \mathcal{J}[\boldsymbol{\kappa}_2, \boldsymbol{\kappa}_2] \coth( \frac{\ell-j}{2} \textrm{arccosh}\left( \zeta \right) )  \right] \right) \nonumber \\ 
&+\exp \left( -    \frac{2m}{\beta}   \frac{1}{\sqrt{A_x \Delta_x}}   \left[ \mathcal{J}[\boldsymbol{\kappa}_1, \boldsymbol{\kappa}_1] + \mathcal{J}[\boldsymbol{\kappa}_2, \boldsymbol{\kappa}_2] \right]\coth \left(   \frac{\ell}{2} \textrm{arccosh}\left(\zeta \right) \right)  \right. \nonumber \\
& \left. \left. - \frac{4m}{\beta} \frac{1}{\sqrt{A_x \Delta_x}}  \mathcal{J}[\boldsymbol{\kappa}_1, \boldsymbol{\kappa}_2]    \frac{\cosh \left(  \left[ \frac{\ell }{2} - j \right]  \textrm{arccosh}\left(\zeta \right) \right) }{\sinh \left( \frac{\ell}{2}  \textrm{arccosh}\left(\zeta \right) \right)}   -  \frac{4m}{\beta} \mathcal{X}[\boldsymbol{\kappa}_1, \boldsymbol{\kappa}_2 ]  \frac{\sinh \left(  \left[ \frac{\ell }{2} - j \right]  \textrm{arccosh}\left(\zeta \right) \right) }{\sinh \left( \frac{\ell}{2}  \textrm{arccosh}\left(\zeta \right) \right)} \right) \right]. 
\label{finaltwosourcedZ}
\end{align}
Here the functionals of the two source terms ($\alpha,\gamma=1,2$) in the exponents of the cycle-dependent contributions are given by:
\begin{align}
 \mathcal{J}[\boldsymbol{\kappa}_\alpha, \boldsymbol{\kappa}_\gamma] = -   \frac{1}{A_x \Delta_x -1 }  &\left[  A_x  \left(  \sum_{n } \frac{\boldsymbol{\kappa}_{\mathbf{\alpha},n} }{\nu_n^2 + \beta x_n} \right) \left(  \sum_{n } \frac{\boldsymbol{\kappa}_{\mathbf{\gamma},n} }{\nu_n^2 + \beta x_n} \right) \right. \nonumber \\
& \left.  - \Delta_x \left( \frac{\beta}{2}    \sum_{n \neq 0}  \frac{ i \nu_n}{   \nu_n^2 + \beta x_n }  \boldsymbol{\kappa}_{\mathbf{\alpha},n} \right) \cdot \left( \frac{\beta}{2}    \sum_{n \neq 0}  \frac{ i \nu_n}{   \nu_n^2 + \beta x_n }  \boldsymbol{\kappa}_{\mathbf{\gamma},n} \right)   \right],  
\end{align}

\begin{align}
 \mathcal{X}[\boldsymbol{\kappa}_1, \boldsymbol{\kappa}_2 ] = - \frac{1}{A_x \Delta_x-1}  & \left[ \sum_{n } \frac{\boldsymbol{\kappa}_{\mathbf{1},n} }{\nu_n^2 + \beta x_n} \cdot \left( \frac{\beta}{2}    \sum_{n \neq 0}  \frac{ i \nu_n}{   \nu_n^2 + \beta x_n }  \boldsymbol{\kappa}_{\mathbf{2},n} \right) \right. \nonumber \\
&  \left.  -      \sum_{n } \frac{\boldsymbol{\kappa}_{\mathbf{2},n} }{\nu_n^2 + \beta x_n} \cdot \left( \frac{\beta}{2}    \sum_{n \neq 0}  \frac{ i \nu_n}{   \nu_n^2 + \beta x_n }  \boldsymbol{\kappa}_{\mathbf{1},n} \right) \right],
\end{align}
and the argument of the cycle-independent exponent in front is given by:
\begin{align}
\tilde{S} [  \boldsymbol{\overline{\kappa}}_2] &=  -\frac{(N-1)m  \beta }{ 2N}   \sum_n \frac{\boldsymbol{\kappa}_{1,n}  \cdot \boldsymbol{\kappa} _{1,-n}+\boldsymbol{\kappa}_{2,n} \cdot \boldsymbol{\kappa}_{2,-n}}{\nu_n^2 + \beta x_n}  -\frac{ m  \beta }{ 2N}  \sum_n \frac{\boldsymbol{\kappa}_{1,n}  \cdot \boldsymbol{\kappa} _{1,-n}+\boldsymbol{\kappa}_{2,n} \cdot \boldsymbol{\kappa}_{2,-n}}{\nu_n^2 + \beta y_n}  \nonumber \\
&+\frac{ m \beta }{ 2N}  \sum_n \frac{2 \boldsymbol{\kappa}_{1,n} \cdot \boldsymbol{\kappa}_{2,-n} }{\nu_n^2 + \beta x_n}  - \frac{ m  \beta }{ 2N}  \sum_n \frac{2 \boldsymbol{\kappa}_{1,n} \cdot \boldsymbol{\kappa}_{2,-n}}{\nu_n^2 + \beta y_n}  - \frac{2m}{\beta}  \left[\mathcal{J}[\boldsymbol{\kappa}_1, \boldsymbol{\kappa}_1] + \mathcal{J}[\boldsymbol{\kappa}_2, \boldsymbol{\kappa}_2]\right].
\end{align}
Expression (\ref{finaltwosourcedZ}) is nothing else than $\expval{\exp \left( m \int_0^\beta d \tau_1 \mathbf{r}_1 (\tau_1) \boldsymbol{\kappa}_1(\tau_1) + m \int_0^\beta d \tau_2 \mathbf{r}_2 (\tau_2) \boldsymbol{\kappa}_2(\tau_2) \right) }$ with the expectation value taken with respect to the unsourced system. 
To obtain the two-point correlation function (\ref{twopoint1}) we set the two source functions equal to respectively $  \mathbf{f}_1(\tau_1) = \frac{i\mathbf{q}}{m} \delta( \tau_1 - \tau)$ and $\mathbf{f}_2(\tau_2) = -\frac{i\mathbf{q}}{m} \delta( \tau_2 - \sigma)$, which leaves the general form of expression (\ref{finaltwosourcedZ}) unchanged except for simplifying the functionals $\mathcal{J}$, $\mathcal{X}$ and $\tilde{S}$ to: 

\begin{align}
 &\mathcal{J}[\mathbf{f}_1, \mathbf{f}_1] =  \frac{q^2 \beta^2}{16  m^2} \frac{1}{A_x \Delta_x -1} \left[ A_x \Delta_x (\tau)^2 - \Delta_x(0) \left( \frac{\beta}{2} \partial_\tau \Delta_x (\tau) \right)^2 \right], \label{Jsimplified} \\
 &\mathcal{J}[\mathbf{f}_2, \mathbf{f}_2] =  \frac{q^2 \beta^2}{16  m^2} \frac{1}{A_x \Delta_x -1} \left[ A_x \Delta_x (\sigma)^2 - \Delta_x \left( \frac{\beta}{2} \partial_\sigma \Delta_x (\sigma) \right)^2 \right] , \\
 &\mathcal{J}[\mathbf{f}_1, \mathbf{f}_2] = - \frac{q^2 \beta^2}{16  m^2} \frac{1}{A_x \Delta_x -1} \left[ A_x \Delta_x (\tau) \Delta_x (\sigma)  - \Delta_x \left( \frac{\beta}{2} \partial_\tau \Delta_x (\tau) \right) \left( \frac{\beta}{2} \partial_\sigma \Delta_x (\sigma)  \right)  \right], \\
 & \mathcal{X}[\mathbf{f}_1, \mathbf{f}_2]= \frac{q^2 \beta^2}{16  m^2} \frac{1}{A_x \Delta_x -1} \left[ \Delta_x (\tau)\left( \frac{\beta}{2} \partial_\sigma \Delta_x (\sigma)  \right)-\Delta_x (\sigma)\left( \frac{\beta}{2} \partial_\tau \Delta_x (\tau)  \right)  \right], \label{Xsimplified}  \\
 & \tilde{S} [  \boldsymbol{\overline{\kappa}}_2] = \frac{q^2 \beta}{4mN} \left[ (N-1) \Delta_x +  \Delta_y  +   \Delta_x(\tau-\sigma) - \Delta_y(\tau-\sigma) \right]  - \frac{2m}{\beta} \left[ \mathcal{J}[\mathbf{f}_1, \mathbf{f}_1] + \mathcal{J}[\mathbf{f}_2, \mathbf{f}_2]\right].
\label{Stildesimplified}
\end{align}
Note that while so far the short hand notation for $\Delta_x=\Delta_x(0)$ was used, in (\ref{Jsimplified})-(\ref{Stildesimplified}) the full time-dependence of $\Delta_x(\tau)$ as defined in (\ref{Dxn}) is invoked. 
Although (\ref{finaltwosourcedZ}) has no closed form expression, the numerical solution mainly requires knowing the factor $\mathbb{Z}(N)$, which is obtained by solving (\ref{recurrence1}) as shown in \cite{paper1}. 
Finally, just as considered in Section (\ref{section2}) for the density, the coupled harmonic oscillator limit of the two-point correlation function can be checked for $\tau=\sigma=0$, and exactly agrees with the results in \cite{Brosens1997b}.

\section{Example application: density and pair correlation functions in an open quantum system}\label{section4}

In this section the expressions derived in Sections (\ref{section2}) and (\ref{section3}) are applied to study the particle density, condensed fraction and two-point correlations of a system of bosons in a model environment. 
We consider $N$ non-interacting bosonic oscillators labeled by the coordinates $\overline{\mathbf{r}}=\lbrace \mathbf{r}_1, \mathbf{r}_2,..., \mathbf{r}_N \rbrace$, coupled to a set of external distinguishable masses labeled by $\overline{\mathbf{Q}}=\lbrace \mathbf{Q}_1, \mathbf{Q}_2,..., \mathbf{Q}_N \rbrace$, where the total system is described by the Lagrangian \cite{paper1}:
\begin{equation}
    L_{\textrm{tot}} = \sum_{i=1}^{N} \left(  \frac{m}{2} \dot{\mathbf{r}}_i^2 +  \frac{m \Omega^2}{2} \mathbf{r}_i^2+\frac{M}{2} \dot{\mathbf{Q}}_i^2 + \frac{MW^2}{2} \left( \mathbf{r}_i - \mathbf{Q}_i \right)^2 \right). 
    \label{Ltot}
\end{equation}
The bosons with mass $m$ are trapped in a harmonic potential with frequency $\Omega$, whereas the external particles with mass $M$ are harmonically coupled to the bosons with a frequency $W$. 
In the rest of this section only equal masses  $M=m$ will be considered, and $W / \Omega$ will be used as the coupling parameter to the environment. 
The external masses are easily integrated out which allows to formulate the behavior of the bosons at the level of action functional (\ref{action1}) where the memory kernels are identified as \cite{paper1}: 

\begin{equation}
x(\tau-\sigma)=y(\tau-\sigma) = \frac{MW^2}{m} \left[  \frac{W^2+ \frac{m}{M} \Omega^2}{W^2} \delta(\tau-\sigma) - \frac{W \cosh( W \left[ |\tau-\sigma|-\beta/2 \right])}{2 \sinh(W \beta /2)}  \right] . \label{memorykernel1}
\end{equation}
Note that contrary to the treatment in \cite{paper1}, every expression studied in this section follows from expectation values, and there is no necessity to explicitly define the external system relative to which the energy would be measured. 
Having obtained the memory kernels (\ref{memorykernel1}) for this system, expressions for $\Delta_x$, $A_x$, $Q_x$ and $\mathbb{Z}(N)$ can be computed and were discussed in \cite{paper1}. 
Since for the two-point correlation functions we will restrict ourselves to equal times $\tau=\sigma=0$, this is sufficient to compute any of the quantities from Section (\ref{section2}) and (\ref{section3}). 
Note that for this particular system we have numerically checked that $\Delta_x A_x >1$ and hence $\zeta>1$, which restricts the use of all inverse hyperbolic functions to their real domain and allows us to drop the absolute value signs.

Before proceeding to the presentation of the results let us consider the one-particle reduced density matrix which simplifies quite a bit in the $x=y$ case as per (\ref{memorykernel1}):
\begin{align}
\rho_1(\mathbf{r}'|\mathbf{r})  =   &  \left( \frac{2m}{\pi \beta} \right)^{d  /2}    \frac{1}{N} \sum_{\ell=1}^{N}  \frac{ \mathbb{Z}(N-\ell) }{\mathbb{Z}(N)  } \left(      \sqrt{\frac{A_x}{\Delta_x}} \tanh \left( \frac{\ell}{2} \textrm{arccosh}\left( \zeta \right) \right) \right)^{d/2} \frac{ 1 }{    \left| 2\sinh( \frac{\ell}{2} \textrm{arccosh}(\zeta) ) \right|^{d}}   \nonumber  \\
& \exp \left( - \frac{m}{\beta} \sqrt{\frac{A_x}{\Delta_x}}  \tanh( \frac{\ell}{2} \textrm{arccosh}(\zeta) )   (\mathbf{r}^2+ \mathbf{r}^{\prime 2})  -  \frac{m}{ \beta}  \sqrt{\frac{A_x}{\Delta_x}}  \frac{1}{\sinh(\ell~\textrm{arccosh}\left(\zeta\right) ) }  (\mathbf{r} - \mathbf{r}')^2 \right) \label{1RDMapplication}.
\end{align}
In general, expr.~(\ref{1RDMapplication}) describes a mixed state at the single particle level due to entanglement with the rest of the system and can be decomposed in terms of a classical ensemble with occupation numbers. 
In Appendix \ref{AppendixC} we show how the spectral decomposition of (\ref{1RDMapplication}) is obtained (in $d=3$):

\begin{equation}
\rho_1(\mathbf{r}'|\mathbf{r}) = \sum_{n_x,n_y,n_z}^{\infty} \lambda_{\mathbf{n}} \psi_{\mathbf{n}}^{*}(\mathbf{r}) \psi_{\mathbf{n}}(\mathbf{r}') ,
\end{equation}
with the effective eigenstates and occupancy numbers given by:
\begin{align}
 &\psi_{\mathbf{n}}(\mathbf{r})  =\mathcal{N}_{\mathbf{n}}   H_{n_x} \left( \sqrt{\alpha } x \right) H_{n_y} \left( \sqrt{\alpha } y \right) H_{n_z} \left( \sqrt{\alpha } z \right) \exp \left( - \alpha  \mathbf{r}^2 \right),  \label{psi_n} \\
 &\lambda_{\mathbf{n}} = \frac{1}{N} \sum_{\ell=1}^{N}  \frac{ \mathbb{Z}(N-\ell) }{\mathbb{Z}(N)  }    e^{- \left(\frac{1}{2}+  n_x+n_y+n_z \right) \ell~ \textrm{arccosh}(\zeta)},
\end{align}
where $H_n$ is a Hermite polynomial, $\mathcal{N}_{\mathbf{n}} $ is the normalization factor of the eigenstate, and $\alpha=\frac{2m}{\beta} \sqrt{\frac{A_x}{\Delta_x}}$.

\begin{figure}[!htb]
\includegraphics[width=0.65\columnwidth]{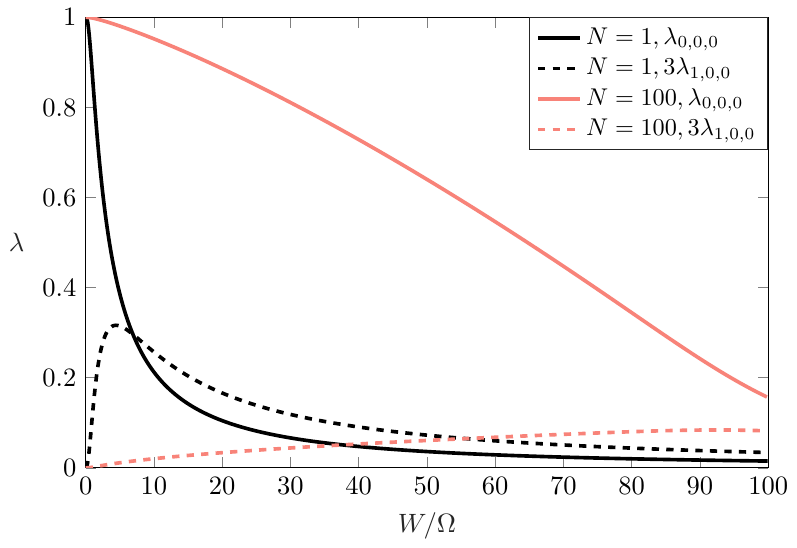}
\caption{The occupation numbers at $T/T_c=0.01$ ($T_c$ defined as the condensation temperature for $N=100$ at $W=0$) of the ground state and the first excited state (counting degeneracy) for respectively $N=1$ and $N=100$, are shown as a function of $W$, the strength of the coupling to the bath of distinguishable particles. } 
\label{Figure2}
\end{figure}

In Fig.~\ref{Figure2} the occupation numbers of the effective ground state and first excited state are compared for respectively $N=100$ bosons and for $N=1$, of which the latter is equivalent to the distinguishable particle case of the system. The results are plotted as a function of the coupling $W$ between the bosons and the bath. 
For numerical purposes the temperature is taken to be finite $T/T_c=0.01$, where $k_B T_c= \hbar \Omega \left( N / \zeta(3)\right)^{1/3} $ for $N=100$. In practice this represents the $T=0$ case as the results have already converged as a function of temperature. 
At $W=0$ and $T=0$ each particle can be described by the same pure $\lambda_{\mathbf{0}}=1$ harmonic oscillator ground-state $\psi_{\mathbf{0}}(\mathbf{r})$ regardless of the particles being distinguishable or not. 
This should not be surprising as in the $T \rightarrow 0$  limit the ground state of $N$ distinguishable non-interacting particles also obeys the bosonic permutation symmetry. 
However, as $W$ is increased, the $N=1$ case rapidly loses its purity as the excited states of the density matrix spike in their occupancy numbers. 
Bosons, on the other hand, retain a macroscopic occupation of the ground state up to far stronger coupling strengths, illustrating how condensation could protect the system from entanglement with the environment.

\begin{figure}[!htb]
\includegraphics[width=0.9\columnwidth]{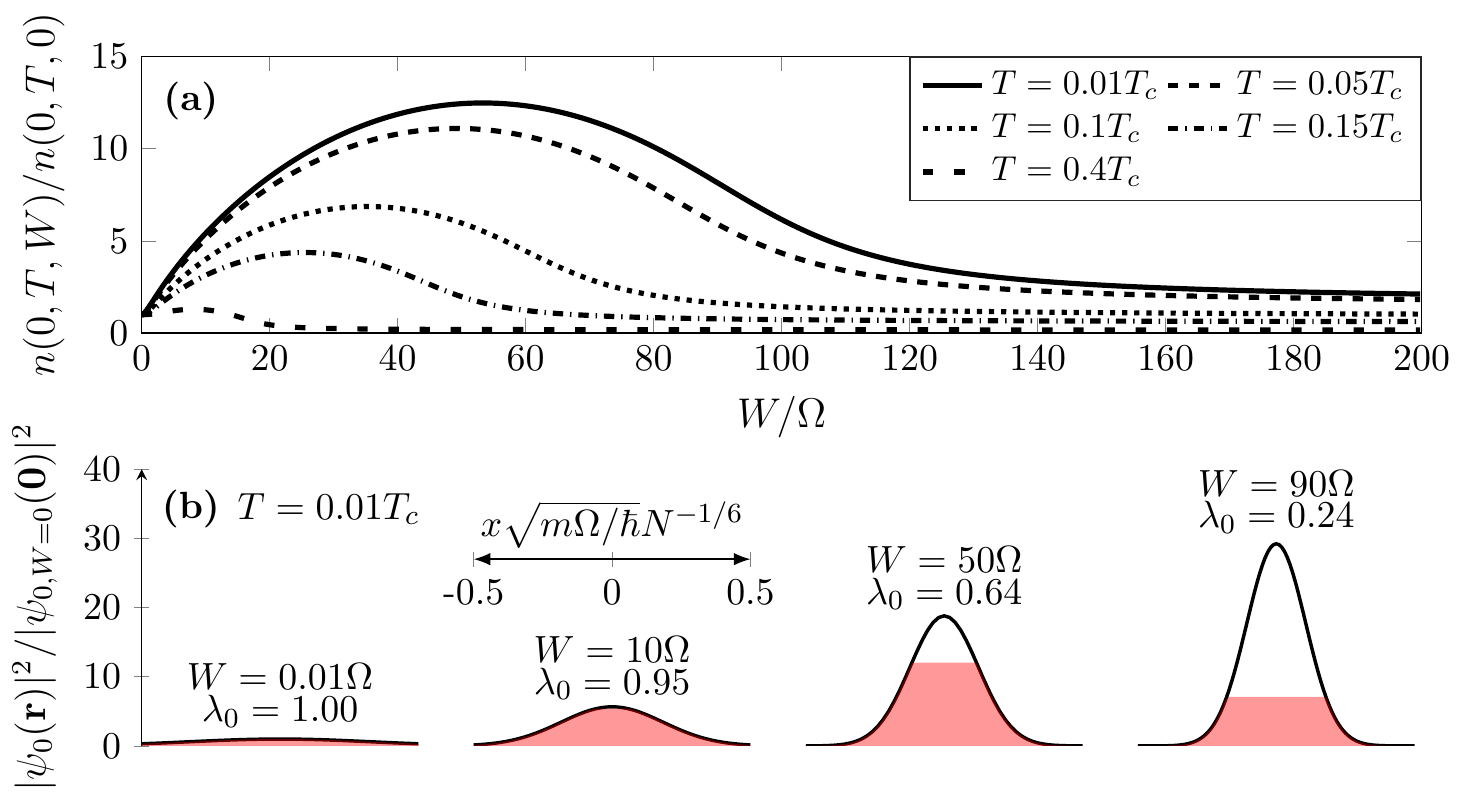}
\caption{Panel (a) presents the central density  of $N=100$ bosons (normalized by its value in the absence of the environment) for a series of low temperatures as a function of $W$. In panel (b) the spatial profile of the ground state is shown at zero temperature (normalized by its maximum value in the absence of an environment) for a set of coupling strengths $W$. The height of the shaded region in the peak relative to the peak height represents the occupation number $\lambda_0$. This also indicates the central density of the ground state fraction as this quantity is proportional to $\lambda_0$.} 
\label{Figure3}
\end{figure}

\begin{figure}[!htb]
\includegraphics[width=0.85\columnwidth]{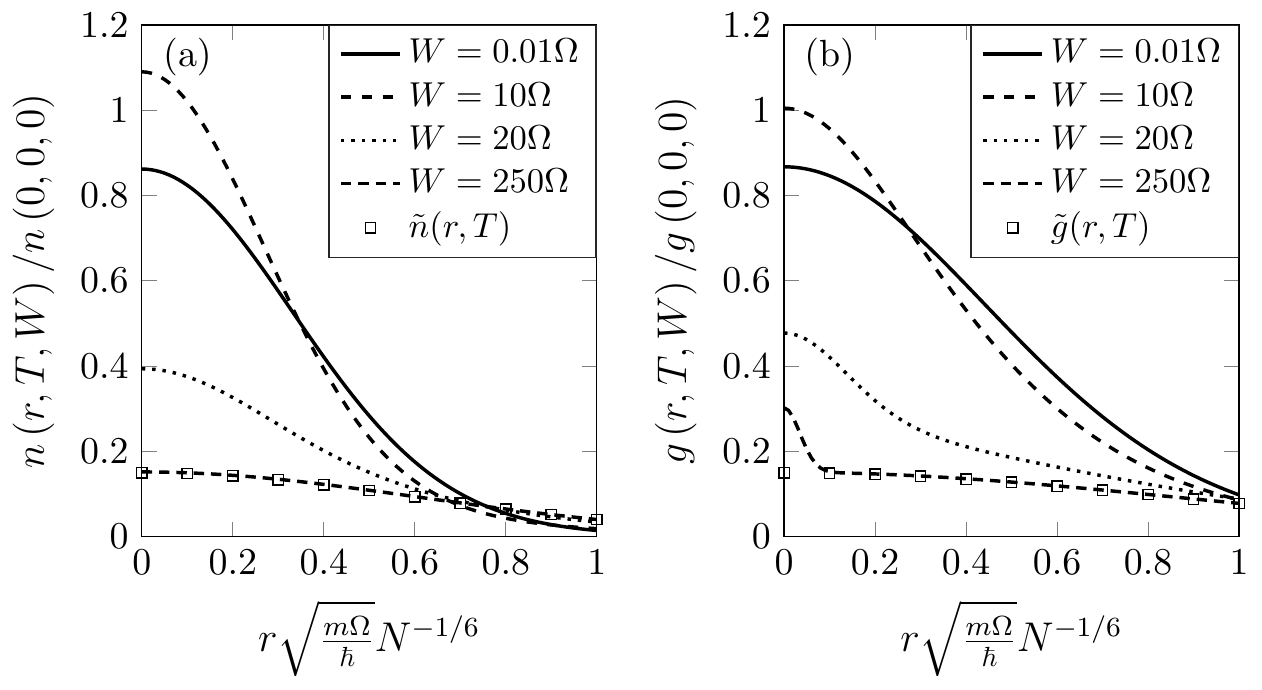}
\caption{The density radial profile (a) and the pair correlation function (b) of $N=100$ bosons for a series of coupling strengths $W$ at a temperature $T=0.4T_c$. The square scatters indicate the asymptotic model results given by (\ref{ntilde}) and (\ref{gtilde}).} 
\label{Figure4}
\end{figure}

The ground state $\psi_{\mathbf{0}}(\mathbf{r})$ gets more sharply peaked when $W$ is increased as can be seen from (\ref{psi_n}), which combined with the behavior of $\lambda_0$ leads to a peculiar behavior of the particle density (\ref{density1}). 
As can be seen in Fig.~\ref{Figure3}a, the central density of the bosonic cloud obtains a non-monotonic behavior as a function of $W$ at low temperatures. 
The origin of this behavior is revealed in Fig.~\ref{Figure3}b: the initial increase in central density as a function of $W$ is due to the compression of the condensate wave function, whereas the subsequent decrease when $W$ is further increased is due to the depletion of the condensate, as depletion overtakes the compression effect on the condensate wave function. 

In the previous discussion we assume that the central density closely mirrors the condensate central density. This is a qualitative argument that neglects the contribution from the excited states compared to the ground state. In contrast to the bosonic case, distinguishable particles do not retain a macroscopic occupation of the ground state and there the central density is determined by the excited states. In this case the above argument will no longer hold which makes the behavior in Fig.~\ref{Figure3} uniquely bosonic. We can also consider the radial profile of the density $n(r)$ shown in Fig.~\ref{Figure4}a where the non-monotonic behavior is clearly visible.

Having obtained an expression for the two-point correlation function (\ref{finaltwosourcedZ}), we can compute the radial pair correlation function representing the average density around each particle as \cite{Brosens1997b}:
\begin{equation}
g(r) =  \frac{N-1}{(2\pi)^d} \int \mathbf{dq} \expval{ e^{ i \mathbf{q}  \cdot \left( \mathbf{r}_1(0) - \mathbf{r}_2(0) \right) }} e^{-i \mathbf{q} \cdot \mathbf{r}}.
\end{equation}
The radial profile of this correlation function is shown in Fig.~\ref{Figure4}b and qualitatively looks nearly identical to the average density profiles. 
This is to be expected since $g(r)$ is still a measure for the particle density, only now conditional to a boson being present at $\mathbf{r}=0$. 
The most noticeable difference is that at strong coupling $W$, the pair correlation exhibits a sharp spike at small distances. 

\begin{figure}[!htb]
\includegraphics[width=0.8\columnwidth]{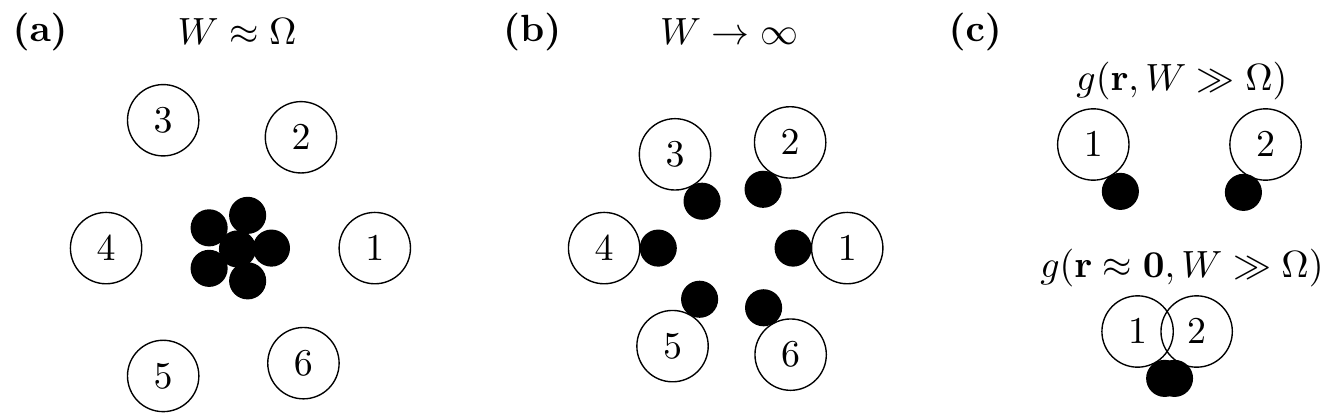}
\caption{A depiction of the asymptotic limit of the model. When $W$ is comparable to the trapping frequency $\Omega$, the bosons can condense, largely remaining indistinguishable (a). When $W$ is increased the particles are glued together and the bosons effectively acquire a distinguishable label (b). Finally note that this picture is less accurate for two-point correlations, even with the distinguishable labels the bosonic nature is retained in the pair-correlation function at short distances (c).} 
\label{Figure5}
\end{figure}

To understand this, it is illustrative to discuss the strong coupling limit of this model. 
Since the external particles in (\ref{Ltot}) are distinguishable, taking $W \rightarrow \infty$ effectively glues them to the bosons giving them a distinguishable label as depicted in Fig.~\ref{Figure5}. 
As a reminder, we are considering the  equal masses case $M=m$, and hence in this limit we should be able to describe the total system as a gas of distinguishable non-interacting composite particles with mass $\tilde{m}=2m$ that are harmonically trapped by a frequency $\tilde{\Omega}=\Omega / \sqrt{2}$. 
The density of such a system is readily written down as the diagonal of the normalized propagator of the harmonic oscillator with mass $\tilde{m}$ and frequency $\tilde{\Omega}$:

\begin{equation}
\tilde{n}(\mathbf{r},T ) = \left( \frac{ \sqrt{2} m \Omega \tanh( \frac{\beta \Omega}{2 \sqrt{2}} )}{\pi} \right)^{d/2} \exp \left( - \sqrt{2} m \Omega \tanh( \frac{\beta \Omega}{2 \sqrt{2}} ) \mathbf{r}^2 \right).
\label{ntilde}
\end{equation}
If the particles are distinguishable and non-interacting the pair correlation function $\tilde{g}$ of this asymptotic model can be computed as:
\begin{align}
\tilde{g} ( \mathbf{r} , T) & = (N-1) \int \mathbf{dr}' \tilde{n}(\mathbf{r} ' ,T ) \tilde{n}(\mathbf{r} ' + \mathbf{r} ,T )  \nonumber \\
&= (N-1) \left( \frac{  m \Omega \tanh( \frac{\beta \Omega}{2 \sqrt{2}} )}{\sqrt{2} \pi} \right)^{d/2} \exp \left( - \frac{1}{\sqrt{2}} m \Omega \tanh( \frac{\beta \Omega}{2 \sqrt{2}} ) \mathbf{r}^2 \right).
\label{gtilde}
\end{align}
These quantities are now plotted alongside the density and pair correlation functions in Fig.~\ref{Figure4}. 
For the density an exact agreement is seen which confirms that the single-particle correlation functions lose all their bosonic properties. 
For the pair correlation function at large distances an exact agreement is found, but at small distances the pair correlation function exhibits a sharp kink which only disappears in the true $W \rightarrow \infty $ limit. 
Therefore we conclude that even when the bosons acquire distinguishable labels, the bosonic properties remain robustly hidden at short distances in the pair correlation functions. 

\section{Conclusion}\label{sec:conclusions}

In this paper we derive the one particle reduced density matrix, the density, and two-point correlation function for a general class of quadratic bosonic systems with retarded interactions in the canonical ensemble. 
As the bosons obtain an effective memory in this description, the commonly used composition properties of path integral propagators no longer hold, and a more general approach to compute the contribution of the permutation cycles is presented. 

This formalism is then applied to a model an open quantum system of identical oscillators coupled to external masses. 
We show how as the coupling strength with the environment is increased, distinguishable particles rapidly get entangled to the external system becoming a highly mixed state, whereas the bosonic case retains its macroscopic occupation of the ground state up to far stronger coupling strengths. 
This gives raise to uniquely bosonic non-monotonic behavior of the particle density as a function of the coupling strength, where at an intermediate coupling the bosons experience maximal trapping strength. 
In the context of the density and pair correlation function the strong coupling limit is discussed, where we show how at sufficiently strong coupling strength even at zero temperature the bosons become distinguishable, while retaining a trace of the bosonic statistics in the short-range part of the pair correlation function.  

The presented results open up the semi-analytic treatment of an entirely new class of action functionals for a finite number of identical particles in the path-integral formalism. 
Retarded interactions have already proven to be a powerful method in variational models for certain types of systems. With the present work, all the prerequisites to formulate a general variational model for identical particles are obtained, which we believe to be a particularly interesting direction to follow up with this approach. 

\begin{acknowledgments}
We gratefully acknowledge fruitful discussions with F. Brosens, S.N. Klimin and M. Houtput. 
T.I. acknowledges the support of 
the Research Foundation-Flanders (FWO-Vlaanderen) 
through the PhD Fellowship Fundamental Research, Project No. 1135521N. 
We also acknowledge financial support from the Research Foundation-Flanders (FWO-Vlaanderen) Grant No. G.0618.20.N, and  from the research council of the University of Antwerp.

\end{acknowledgments}

\appendix 

\section{Calculation of the open cycle}\label{appendixA}

In this appendix an expression for $O_\ell(\mathbf{k})$ given in (\ref{O1}) is computed. 
Since $O_\ell(\mathbf{k})$ factorizes in terms of its dimensional components it is sufficient to do the derivation in $d=1$:

\begin{equation}
O_\ell (k) =  \int d z_2  ... \int d z_\ell ~  K[x ] ( z', \beta | z_\ell, 0 ) ... K[x ] ( z_3, \beta | z_2 , 0 )   K[x ] ( z_2, \beta | z , 0 ) e^{ - \frac{i\mathbf{k}  }{N}   \cdot  \sum_{j=2}^{\ell} z_j} \label{O1appendix1}.
\end{equation}
After substituting the expressions for the propagators (\ref{propagatorfactorization1}) and performing the Gaussian integral we can write:
\begin{align}
O_\ell (k) = \mathcal{A}^{\ell} \exp \left[ - (a+b) \left( z'^2+ z^2 \right) \right]   \sqrt{\frac{\pi^{\ell-1}}{\det(\mathcal{T})} } \exp \left( \frac{1}{4} \left(c \mathbf{u}^T + \boldsymbol{\alpha}^T \right) \mathcal{T}^{-1} \left(c \mathbf{u} + \boldsymbol{\alpha} \right) \right),
\label{O1appendix2}
\end{align}
where precisely as in \cite{paper1} we define the shorthand notations $a= \frac{m}{2\beta} A_x$, $b=\frac{m}{2\beta} \frac{1}{\Delta_x}$. In addition we define the following vectors in (\ref{O1appendix2}): $\boldsymbol{\alpha}^T = ( 2(a-b)z ,0, ..., 0, 2(a-b)z'  )$,
$\mathbf{z}^T= ( z_2, z_3, ..., z_\ell$), and $\mathbf{u}^T= (1,1,...,1)$ with $c=-i k/N$.
The main difference with the open cycles computed in \cite{paper1} is that now the central object is the $\ell-1 \times \ell-1$ dimensional tridiagonal Toeplitz matrix:
\begin{equation}
\mathcal{T}= \begin{pmatrix}
2(a+b) & (b-a) & 0 & ... & 0 \\
(b-a) & 2(a+b) & (b-a)   & ...& ... \\
0     &  (b-a) & 2(a+b) & ... & 0\\
...   & ...    & ... & ... & (b-a) \\
 0 & ... & 0 & (b-a) &  2(a+b) 
\end{pmatrix} ,
\end{equation}
which clearly loses the cyclic symmetry of the circulant matrices that appear in calculations of closed cycles.

The $j=\lbrace 1,...,\ell-1 \rbrace$ eigenvalues of the matrix $\mathcal{T}$ are similar to those of the corresponding ciruclant matrix, but have a longer period in the cosine \cite{Noschese}:
\begin{equation}
\lambda_j = 2(a+b) + 2(b-a) \cos( \frac{j \pi}{\ell} ).
\end{equation}  
The determinant of this matrix is then given by \cite{Fonesca2020}:
\begin{equation}
\det(\mathcal{T})= (a-b)^{\ell-1}   U_{\ell-1} \left( \frac{a+b}{a-b} \right) ,
\label{detT}
\end{equation}
where $U_{\ell-1} \left( \frac{a+b}{a-b} \right)$ is the Chebyshev polynomial of the second kind. 
Here, we define $\zeta=\frac{a+b}{a-b} $ and restrict ourselves to strictly positive $a$ and $b$ with $a \neq b$. 
If $a-b>0$ then $\zeta>1$ and $U_{\ell-1}(\zeta)$ is strictly positive. If $a-b<0$ then then $\zeta<-1$ and $U_{\ell-1}(\zeta)$ can become negative for odd $\ell-1$, which gets compensated by the additional negative sign from $(a-b)^{\ell-1}$. 
Therefore $\det(\mathcal{T})$ is always positive and well-defined.

For $|\zeta|>1$ Chebyshev polynomials of the second kind can also be written as (with any choice of approaching the branch cut of the $\textrm{arccosh}$ function):
\begin{equation}
U_{\ell-1}(\zeta) = \frac{\sinh(  \ell~\textrm{arccosh}\left(\zeta \right) ) }{\sinh(  \textrm{arccosh}\left(\zeta\right) )  } .
\label{Chebyshev}
\end{equation}
This yields the factor in front of (\ref{O1appendix2}), which leaves to find the quadratic form of the inverse in the exponent. 
The inverse elements of matrix $\mathcal{T}$ are given by \cite{Fonseca,Fonesca2020} (where we have taken out an additional minus sign out of the Chebyshev polynomials):
\begin{align*}
&T^{-1}_{ij}=  \sigma_{ij}= \frac{1}{a-b} \frac{U_{i-1}\left( \zeta \right) U_{\ell-1-j} \left( \zeta  \right)}{U_{\ell-1} \left( \zeta \right)}  \hspace{10pt} \textrm{if}  \hspace{10pt} i \leq j ,\\
&T^{-1}_{ij}=  \frac{1}{a-b} \frac{U_{j-1}\left( \zeta \right) U_{\ell-1-i} \left(\zeta \right)}{U_{\ell-1} \left( \zeta\right)}  \hspace{10pt} \textrm{if}  \hspace{10pt} i > j.
\end{align*}
Since we have assumed $a>0$ and $b>0$ we can use the results of \cite{Fonseca} to write:
\begin{equation}
\left( \mathcal{T}^{-1} \mathbf{u} \right)_i = s_i= \frac{1+(b-a)(\sigma_{1,i} + \sigma_{1,\ell-i}) }{4b}  \hspace{10pt} \textrm{and} \hspace{10pt} \mathbf{u}^T   \mathcal{T}^{-1}  \mathbf{u} = \frac{(\ell-1)+2(b-a) s_1 }{4b}.
\end{equation}
This can now be used to compute all the necessary terms in the quadratic form in the exponent of (\ref{O1appendix2}): 
\begin{align}
&\mathbf{u}^T   \mathcal{T}^{-1}  \mathbf{u} =   \frac{1}{4b} \left[  \ell - \sqrt{\frac{a}{b}}  \tanh( \frac{\ell}{2} \textrm{arccosh}(\zeta) )  \right] ,\label{part1} \\
&\boldsymbol{\alpha}^T \mathcal{T}^{-1} \mathbf{u} = \left[  \sqrt{\frac{a}{b}}  \tanh( \frac{\ell}{2} \textrm{arccosh}(\zeta) )  -1 \right](z+z'), \label{part2} \\
& \boldsymbol{\alpha}^T \mathcal{T}^{-1} \boldsymbol{\alpha} = \left( 4(a+b) - 8 \sqrt{ab}  \tanh( \frac{\ell}{2} \textrm{arccosh}(\zeta) )  \right) (z^2+ z^{\prime 2}) -  \frac{4(a-b)}{U_{\ell-1}(\zeta)} (z - z')^2,  \label{part3}
\end{align}
where we have used the easily proven identity:
\begin{equation}
\frac{U_{\ell-2}(\zeta)}{U_{\ell-1}(\zeta)} + \frac{1}{U_{\ell-1}(\zeta)} = \frac{a+b}{a-b} -  \frac{2\sqrt{ab} }{a-b} \tanh( \frac{\ell}{2} \textrm{arccosh}(\zeta) ).
\end{equation}
Substituting both (\ref{detT}) and (\ref{part1}-\ref{part3}) into expression (\ref{O1appendix2}) finally yields:
\begin{align}
O_\ell (k) = & \mathcal{A}^{\ell}   \sqrt{\frac{\pi^{\ell-1}}{ (a-b)^{\ell-1} U_{\ell-1}( \zeta) } }  \exp \left( - \frac{k^2}{16N^2 b} \left[  \ell - \sqrt{\frac{a}{b}}  \tanh( \frac{\ell}{2} \textrm{arccosh}(\zeta) )  \right]   \right. \nonumber \\
&  - \frac{ik}{2N} \left[  \sqrt{\frac{a}{b}}  \tanh( \frac{\ell}{2} \textrm{arccosh}(\zeta) )  -1 \right](z+z') \nonumber        \\
&  \left. - 2 \sqrt{ab}  \tanh( \frac{\ell}{2} \textrm{arccosh}(\zeta) )   (z^2+ z^{\prime 2}) -  \frac{(a-b)}{U_{\ell-1}(\zeta)} (z - z')^2  \right) . \label{O1appendix3} 
\end{align}
After substitution of $a$, $b$ and $\mathcal{A}$, and after generalization to $d=3$ this yields exactly expression (\ref{O2}).

\section{Calculation of the closed correlation cycles}\label{appendixB}

Let us start this section by computing the two-point correlation cycle (\ref{chi1}), which should also yield the one-point cycle (\ref{H1}) for $\boldsymbol{\kappa}_2=0$. 
Quite similarly to Appendix \ref{appendixA} the computation is done in $d=1$:
\begin{align}
\chi_{\ell}(k,j) = \int dz_1  ... \int dz_\ell ~ & e^{\tilde{a}_1(z_2+z_1) +\tilde{b}_1(z_2-z_1)} K[x ] ( z_1, \beta | z_\ell, 0 ) ...  K[x ] ( z_2, \beta | z_1 0 ) e^{ - \frac{i k }{N}   \cdot  \sum_{j=1}^{\ell} z_j} \nonumber \\
&e^{\tilde{a}_2(z_{j+2}+z_{j+1}) +\tilde{b}_2(z_{j+2}-z_{j+1})} \label{Chi1Appendix},
\end{align}
and will be generalized at the end. Note that for $j=\ell-1$ some care should be taken as $z_{j+2}$ loops back to $z_1$. 
In the derivation below we will implicitly assume $1 < j < \ell-1$, but each step can be readily checked to hold for the boundary cases as well and the obtained result holds for any $1 \leq j \leq \ell-1$. 
We can use the same notation for $c$, $\mathbf{u}$ and $\mathcal{A}$ as in Appendix \ref{appendixA} (but in $\ell$ dimensions) and define the vector:
\begin{equation}
\boldsymbol{w}^T_j= \left( \tilde{a}_1 - \tilde{b}_1, \tilde{a}_1 +\tilde{b}_1,0,...,\tilde{a}_2 - \tilde{b}_2, \tilde{a}_2 + \tilde{b}_2,0,...   \right),
\end{equation}  
which has zeroes everywhere except for the positions: $1,2,j+1,j+2$. 
After substitution of the propagators, the Gaussian integral in (\ref{Chi1Appendix}) is readily performed:
\begin{align}
\chi_{\ell}(k,j)= \mathcal{A}^{\ell}   \sqrt{\frac{\pi^{\ell}}{\det(\mathcal{C})} } \exp \left( \frac{1}{4} \left(c \mathbf{u}^T + \boldsymbol{w}_j^T \right) \mathcal{C}^{-1} \left(c \mathbf{u} + \boldsymbol{w}_j \right) \right).
\label{Chi1Appendix2}
\end{align}

The cycle considered here is closed and hence just like in \cite{paper1} the central object appearing is the $\ell \times \ell$ dimensional three-circulant matrix $\mathcal{C}$ that is defined by a periodic shifting of the first row:
\begin{equation}
\mathcal{C}=\textrm{circ} \left(2(a+b),(b-a),0,...,(b-a) \right)= \begin{pmatrix}
2(a+b) & (b-a) & 0 & ... & (b-a) \\
(b-a) & 2(a+b) & (b-a)   & ...& ... \\
0     &  (b-a) & 2(a+b) & ... & 0\\
...   & ...    & ... & ...  & (b-a) \\
(b-a) & ... & 0 & (b-a) & 2(a+b) 
\end{pmatrix} ,
\label{circulantC}
\end{equation}
where the same shorthand notation for $a$ and $b$ is used as in Appendix \ref{appendixA}. 
The properties of this matrix are discussed in \cite{CirculantGray} and the determinant is given by \cite{paper1}:
\begin{equation}
\det( \mathcal{C} ) =  4 \left[ (a-b) \right] ^\ell  \sinh( \frac{\ell}{2}  \textrm{arccosh}\left( \frac{a+b}{a-b} \right) )^2. 
\label{determinant2}
\end{equation}
To find $\chi_{\ell}(k,j)$ therefore only the quadratic form of the inverse matrix in the exponent (\ref{Chi1Appendix2}) has to be computed. 
Using circulant matrix properties \cite{CirculantGray} we can write $\mathcal{C}^{-1}= Q D^{-1} Q^*$. 
Here, $D$ is the diagonal matrix of eigenvalues $\lambda_j= 2(a+b)+2(b-a)\cos(\frac{2\pi j}{\ell})$ and $Q$ is the matrix that has the the eigenvectors $\mathbf{y}^T_j=  \left( \rho^{j0}, \rho^{j1},..., \rho^{j (\ell-1)} \right)$ as columns, where $\rho=\exp\left( \frac{2 \pi i}{\ell} \right)$. 
For expressions appearing in the first three terms of the quadratic form in (\ref{Chi1Appendix2}) this readily yields:
\begin{equation}
\mathbf{u}^T \mathcal{C}^{-1} \mathbf{u}  = \frac{\ell}{4b} \hspace{20pt} \textrm{and} \hspace{20pt} \boldsymbol{w}_j^T \mathcal{C}^{-1} \mathbf{u}=  \frac{1}{2b} \left( \tilde{a}_1 + \tilde{a}_2 \right).
\label{parts123}
\end{equation}
The computation of the last part $\boldsymbol{w}_j^T \mathcal{C}^{-1}  \boldsymbol{w}_j=\boldsymbol{w}_j^T Q D^{-1} Q^*  \boldsymbol{w}_j$ is slightly more involved. First we start by explicitly writing:
\begin{equation}
(\boldsymbol{w}_j^T Q)_m =(Q^* \boldsymbol{w}_j )^*_m = \frac{1}{\sqrt{\ell} } \left[ \left(  \tilde{a}_1 - \tilde{b}_1 \right) + \rho^{m} \left(  \tilde{a}_1+ \tilde{b}_1 \right)+\rho^{mj} \left(  \tilde{a}_2 - \tilde{b}_2 \right)+\rho^{m(j+1)} \left(  \tilde{a}_2 + \tilde{b}_2 \right) \right],
\end{equation}
from which follows:
\begin{align}
\boldsymbol{w}_j^T \mathcal{C}^{-1}  \boldsymbol{w}_j &= \left( \tilde{a}_1^2+ \tilde{b}_1^2+\tilde{a}_2^2+\tilde{b}_2^2 \right) D_\ell(0) + \left(\tilde{a}_1^2-\tilde{b}_1^2 + \tilde{a}_2^2-\tilde{b}_2^2\right) D_\ell(1)  + 2\left( \tilde{a}_1 \tilde{a}_2 + \tilde{b}_1 \tilde{b}_2  \right) D_\ell(j) \nonumber \\
& + ( \tilde{a}_1 - \tilde{b}_1) ( \tilde{a}_2 + \tilde{b}_2) D_{\ell}(j+1)+ (\tilde{a}_1 + \tilde{b}_1)(\tilde{a}_2-\tilde{b}_2) D_\ell(j-1),
\label{part4}
\end{align}
where for any $0 \leq  n \leq  \ell $:
\begin{equation}
D_\ell(n)=\frac{1}{\ell} \sum_{m=0}^{\ell-1} \frac{ \rho^{mn} +\rho^{-mn}}{\lambda_m}= \frac{2}{\ell} \sum_{m=0}^{\ell-1} \frac{   1}{\lambda_m \rho^{-mn}} .
\label{DnAppendixC}
\end{equation}
The reasoning below to compute $D_\ell(n)$ strongly relies on several properties of circulant matrices discussed in \cite{searle}. 
For any general circulant matrix $\mathcal{M}=\textrm{circ} \left( c_0 , c_1 , ..., c_{\ell-1} \right)$ with eigenvalues given by \cite{CirculantGray}:
\begin{equation}
\lambda_m = \sum_{j=0}^{\ell-1} c_j \rho^{mj} ,
\end{equation}
it is not difficult to see that the factors appearing in the denominator of (\ref{DnAppendixC}) can be written as:
\begin{equation}
\lambda_m \rho^{-mn} = \sum_{j=0}^{\ell-1} c_j \rho^{m(j-n)} = c_n \rho^0  + c_{n+1} \rho^{m} + c_{n+2} \rho^{2m} + ...
\end{equation}
This is nothing else than the set of eigenvalues of a circulant matrix of which the initial row has been shifted by $n$ to the left $P_{-n} \mathcal{M}= \textrm{circ} \left( c_n, c_{n+1},... , c_{\ell-1}, c_0, ... \right)$, where $P_{n}$ is defined as the circulant matrix that shifts all the rows of $\mathcal{M}$ by one column to the right in the notation of \cite{searle}. 
Since circulants commute under multiplications it follows that $\left( P_{-n} \mathcal{M} \right)^{-1} = P_n \mathcal{M}^{-1}$, which allows to write the summation (\ref{DnAppendixC}) as:
\begin{equation}
D_n= \frac{2}{\ell} \textrm{Tr} \left( P_n \mathcal{C}^{-1} \right).
\end{equation}
The inverse of a three-circulant (\ref{circulantC}) is computed in \cite{searle}:
\begin{equation}
\mathcal{C}^{-1}= \textrm{circ}\left(d_0, d_1,...,d_{\ell-1} \right)
\end{equation}
from which follows for $0<n \leq \ell$:
\begin{equation}
D_\ell(n) = 2 d_{\ell-n}   \hspace{20pt} \textrm{and} \hspace{20pt} D_\ell(0)= 2 d_0.
\end{equation}
Following previous assumptions that the coefficients of the circulant matrix (\ref{circulantC}) $a$ and $b$ are strictly positive with $a\neq b$, from \cite{searle} follows the following result after some substitutions:
\begin{equation}
d_{\ell-n}  = \frac{1}{ 4\sqrt{ab}}  \frac{ \left(\frac{\sqrt{a}+\sqrt{b}}{\sqrt{a}-\sqrt{b}} \right)^{\ell/2-n}+ \left(\frac{\sqrt{a}-\sqrt{b}}{\sqrt{a}+\sqrt{b}} \right)^{\ell/2-n}}{\left(\frac{\sqrt{a}+\sqrt{b}}{\sqrt{a}-\sqrt{b}} \right)^{\ell/2}- \left(\frac{\sqrt{a}-\sqrt{b}}{\sqrt{a}+\sqrt{b}} \right)^{\ell/2}}.
\end{equation}
Having already cast all the expressions into a goniometric form in \cite{paper1} and Appendix \ref{appendixA}, we can do the same here and write:
\begin{equation}
D_\ell(n)= \frac{1}{ 2\sqrt{ab}}  \frac{\cosh \left[ \left(\frac{\ell}{2}-n \right) \textrm{arccosh}\left(\zeta \right) \right]}{\sinh \left[ \frac{\ell}{2} ~ \textrm{arccosh}\left(\zeta \right) \right]},
\end{equation}
with $\zeta=\frac{a+b}{a-b}$. After substituting (\ref{determinant2}), (\ref{parts123}) and (\ref{part4}) into (\ref{Chi1Appendix2}), we can write: 
\begin{align}
\chi_{\ell}(k,j)= &\mathcal{A}^{\ell}   \sqrt{\frac{\pi^{\ell}}{4  (a-b)  ^\ell  \sinh \left[ \frac{\ell}{2}  \textrm{arccosh}\left( \zeta \right) \right]^2 }}  \exp \left( \frac{-\ell k^2}{16N^2b} -  \frac{i k}{4N b} \left( \tilde{a}_1 + \tilde{a}_2 \right)   \right. \nonumber \\
&+ \frac{1}{4} \left( \tilde{a}_1^2+ \tilde{b}_1^2+\tilde{a}_2^2+\tilde{b}_2^2 \right) D_\ell(0) + \frac{1}{4} \left(\tilde{a}_1^2-\tilde{b}_1^2 + \tilde{a}_2^2-\tilde{b}_2^2\right) D_\ell(1)  + \frac{1}{2} \left( \tilde{a}_1 \tilde{a}_2 + \tilde{b}_1 \tilde{b}_2  \right) D_\ell(j) \nonumber \\
& \left. + \frac{1}{4} ( \tilde{a}_1 - \tilde{b}_1) ( \tilde{a}_2 + \tilde{b}_2) D_\ell(j+1)+ \frac{1}{4}  (\tilde{a}_1 + \tilde{b}_1)(\tilde{a}_2-\tilde{b}_2) D_\ell(j-1) \right).
\label{ChiAppendix3}
\end{align}
To obtain the one-point cycle one has just to substitute $\tilde{a}_2=\tilde{b}_2=0$ and find:
\begin{align}
H^{(1)}_\ell  (k)= &\mathcal{A}^{\ell}   \sqrt{\frac{\pi^{\ell}}{4  (a-b)  ^\ell  \sinh \left[ \frac{\ell}{2}  \textrm{arccosh}\left( \zeta \right) \right]^2 }}  \nonumber \\
&\exp \left( - \frac{ \ell k^2}{16 N^2 b} -  \frac{ i k \tilde{a}_1}{4Nb }  +  \frac{  \tilde{a}_1^2 a - \tilde{b}_1^2 b  }{a-b} \frac{1}{4 \sqrt{ab}} \coth( \frac{\ell}{2} \textrm{arccosh}\left( \zeta \right) ) - \frac{1}{4} \frac{\tilde{a}^2_1-\tilde{b}^2_1}{a-b} \right).
\label{H1appendix3}
\end{align}
The generalization to $d=3$ yields the results presented in the main text in (\ref{Hresultmain}) and (\ref{Chiresultmain}).
\section{Spectral decomposition of the one-particle reduced density matrix}\label{AppendixC}
The one-particle reduced density matrix (\ref{1RDMapplication}) can be written as a summation over Gaussian states:
\begin{equation}
\rho_1(\mathbf{r}'|\mathbf{r}) = \frac{1}{N} \sum_{\ell=1}^{N} g^{(\ell)}(\mathbf{r}'|\mathbf{r}),
\label{rho1appendix}
\end{equation}
and after rewriting the exponents in (\ref{1RDMapplication}) the terms can be written as:
\begin{align}
  g^{(\ell)}(\mathbf{r}'|\mathbf{r}) = C  \left( \frac{\gamma-\eta}{\pi} \right)^{d/2} \exp \left( - \frac{\gamma_\ell }{2}   (\mathbf{r}^2+ \mathbf{r}^{\prime 2})   +  \eta_\ell  \mathbf{r} \cdot \mathbf{r}'  \right).
\end{align}
with:
\begin{align}
& C_\ell =  \frac{ \mathbb{Z}(N-\ell) }{\mathbb{Z}(N)  }  \frac{ 1 }{    \left| 2\sinh( \frac{\ell}{2} \textrm{arccosh}(\zeta) ) \right|^{d}}  \\
&\gamma_\ell  = \frac{2m}{\beta} \sqrt{\frac{A_x}{\Delta_x}}  \coth( \ell~ \textrm{arccosh}(\zeta) ),  \\
& \eta_\ell = \frac{2m}{ \beta}  \sqrt{\frac{A_x}{\Delta_x}}  \frac{1}{\sinh(\ell~\textrm{arccosh}\left(\zeta\right) ) } .
\end{align}
Here, the Gaussian states were suggestively written in this form to use the results from \cite{Srednicki}. 
This allows write down the solution to the Gaussian eigenvalue problem in for $\mathbf{n}= (n_x, n_y,n_z)$ (in $d=3$):

\begin{equation}
\int  \mathbf{dr}'    g^{(\ell)}(\mathbf{r} |\mathbf{r}') \psi_\mathbf{n}^{(\ell)} (\mathbf{r}') = \lambda^{(\ell)}_\mathbf{n}(\mathbf{r} ) \psi_\mathbf{n}^{(\ell)}
\end{equation}
as:
\begin{align}
&\lambda_\mathbf{n}^{(\ell)}=  C_\ell  \left( 1 - \xi_\ell  \right)^d \xi_\ell^{n_x+n_y+n_z } ,\\
& \psi_n(\mathbf{r})^{(\ell)} =  \mathcal{N}_{\mathbf{n}} H_{n_x} \left( \sqrt{\alpha_\ell } x \right) H_{n_y} \left( \sqrt{\alpha_\ell } y \right) H_{n_z} \left( \sqrt{\alpha_\ell } z \right) \exp \left( - \alpha_\ell  \mathbf{r}^2 /2 \right) ,
\end{align}
with $H_n$ a Hermite polynomial, $\alpha_\ell = \left(  \gamma_\ell ^2 - \eta_\ell ^2 \right)^{1/2}$, $\xi_\ell= \frac{\eta_\ell }{\gamma_\ell  +\alpha_\ell }$ \cite{Srednicki}, and the normalization factor $\mathcal{N}_\mathbf{n}=\left( \frac{1}{2^{n_x+n_y+n_z} n_x! n_y! n_z!} \right)^{1/2} \left( \frac{  \alpha_\ell}{\pi} \right)^{d/4}$. 
Remarkably, the $\ell$-dependence in the coefficient $\alpha_\ell$ drops out:
\begin{equation}
\alpha= \frac{2m}{\beta} \sqrt{\frac{A_x}{\Delta_x}} \left(\coth \left[ \ell~ \textrm{arccosh}(\zeta) \right]^2- \frac{1}{\sinh \left[ \ell~\textrm{arccosh}\left(\zeta\right) \right]^2 }\right) ^{1/2} = \frac{2m}{\beta} \sqrt{\frac{A_x}{\Delta_x}}.
\end{equation}
This implies that every Gaussian state $  g^{(\ell)}(\mathbf{r}'|\mathbf{r}) $ has the same set of eigenstates, which are also immediately the eigenstates of  (\ref{rho1appendix}):
\begin{equation}
 \psi_{\mathbf{n}}(\mathbf{r})  =  \mathcal{N}_{\mathbf{n}} H_{n_x} \left( \sqrt{\alpha } x \right) H_{n_y} \left( \sqrt{\alpha } y \right) H_{n_z} \left( \sqrt{\alpha } z \right) \exp \left( - \alpha  \mathbf{r}^2 \right).
\end{equation}
The factor $\xi_\ell$ does remain $\ell$-dependent:
\begin{equation}
\xi_\ell=  \frac{1}{\sinh \left[ \ell~\textrm{arccosh}\left(\zeta\right) \right] } \frac{1 }{1+ \coth \left[ \ell~ \textrm{arccosh}(\zeta) \right]} = e^{- \ell~ \textrm{arccosh}(\zeta)},
\end{equation}
and hence the eigenvalue of the full density matrix (\ref{rho1appendix}) corresponding to state $\psi_{\mathbf{n}}(\mathbf{r})$ is given by:
\begin{equation}
\lambda_{\mathbf{n}} = \frac{1}{N} \sum_{\ell=1}^{N}  \frac{ \mathbb{Z}(N-\ell) }{\mathbb{Z}(N)  }  \frac{ 1 }{    \left| 2\sinh \left[ \frac{\ell}{2} \textrm{arccosh}(\zeta) \right] \right|^{d}} \left( 1 -e^{- \ell~ \textrm{arccosh}(\zeta)} \right)^{d} e^{- (n_x+n_y+n_z)\ell~ \textrm{arccosh}(\zeta)}.
\end{equation}
In the case that $\zeta >1$ this can be simplified even further:
\begin{equation}
\lambda_{\mathbf{n}} = \frac{1}{N} \sum_{\ell=1}^{N}  \frac{ \mathbb{Z}(N-\ell) }{\mathbb{Z}(N)  }    e^{- \left(\frac{1}{2}+  n_x+n_y+n_z \right) \ell~ \textrm{arccosh}(\zeta)}.
\end{equation}

\end{document}